\documentclass[aps,llncs,preprint]{revtex4-1}
\usepackage{graphicx}
\usepackage{amsmath}
\usepackage{float}
\usepackage{enumitem}
\usepackage{subcaption}
\usepackage{color}
\DeclareUnicodeCharacter{2212}{-}
\begin{document}
\title{Tip-induced Superconductivity}

\author{Sandeep Howlader}
\author{Goutam Sheet}
\email{goutam@iisermohali.ac.in}
\affiliation{Department of Physical Sciences, 
Indian Institute of Science Education and Research Mohali, 81, Knowledge city, S.A.S. Nagar, Manauli 140306, Punjab, India}

\begin{abstract}

It is widely believed that topological superconductivity, a hitherto elusive phase of quantum matter, can be achieved by inducing superconductivity in topological materials. In search of such topological superconductors, certain topological insulators (like, Bi$_2$Se$_3$) were successfully turned into superconductors by metal-ion (Cu, Pd, Sr, Nb etc. ) intercalation. Superconductivity could be induced in topological materials through applying pressure as well. for example, a pressure-induced superconducting phase was found in the topological insulator Bi$_2$Se$_3$. However, in all such cases, no conclusive signature of topological superconductivity was found. In this review, we will discuss about another novel way of inducing superconductivity in a non-superconducting topological material -- by creating a mesoscopic interface on the material with a non-superconducting, normal metallic tip where the mesoscopic interface becomes superconducting. Such a phase is now known as a tip-induced superconducting (TISC) phase. This was first seen in 2014 on Cd$_3$As$_2$ at IISER Mohali, India. Following that, a large number of other topological materials were shown to display TISC. Since the TISC phase emerges only at a confined region under a mesoscopic point contact, traditional bulk tools for characterizing  superconductivity cannot be employed to detect/confirm such a phase. On the other hand, such a point contact geometry is ideal for probing the possible existence of a temperature and magnetic field dependent superconducting energy gap and a temperature and magnetic field dependent critical current. We will review the details of the experimental signatures that can be used to prove the existence of superconductivity even when the ``text-book" tests for detecting superconductivity cannot be performed. Then, we will review different systems where a TISC phase could be realized. 

\end{abstract}
\maketitle
\textbf{Contents\\
\\}
\textbf{I. INTRODUCTION\\
\\
II.POINT CONTACT SPECTROSCOPY
\begin{enumerate}[label= \Alph*.]
 \item{ Fabrication of point contacts}  
\item{Electrical Measurements}          
\item{Different regimes of point-contact transport}
\end{enumerate}
III. POINT CONTACT ANDREEV REFLECTION(PCAR)
\begin{enumerate}[label= \Alph*.]
\item{Andreev reflection as a spectroscopic tool} 
       \begin{enumerate}[label=\arabic*.]
          \item{ Blonder-Tinkham-Klapwijk theory}
           \item{ Spin polarized BTK theory}
            \item{ Broadening of DOS}
         \end{enumerate}
\item{ The role of critical current}
 \end{enumerate}
IV. TIP-INDUCED SUPERCONDUCTIVITY(TISC)
\begin{enumerate}[label= \Alph*.] 
 \item{How to detect TISC with confidence?}
\item{Examples of TISC}
 \begin{enumerate}[label=\arabic*.]
    \item{TISC in Dirac semimetals}
    \item{TISC in Weyl semimetals}
 \item{TISC in Nodal semimetals}
 \item{TISC in topological crystalline insulators\\} 
 \end{enumerate} 
 \end{enumerate} }

\section{Introduction}
Superconductivity is a phenomenon where a material loses its resistance completely below a certain critical temperature. Superconductivity was discovered in 1911 when mercury showed immeasurably small resistance below 4.2 K. Later on, in 1933,  perfect diamagnetism was also shown to be a generic property of superconductivity. After mercury, many other elemental metals were shown to superconduct at liquid helium temperatures. Niobium (Nb) showed the highest $T_c$ $\sim$ 9.2 K among the elemental metals. Subsequently, superconductivity was discovered in binary alloys and compounds such as SbSn, $Bi_5Tl_3$ etc. There was a brief hiatus in the superconductivity research for most experimental groups during the second world war. However, in Germany, a new superconductor NbN was discovered for which the $T_c$ was found to be 15 K -- this was the first superconductor to surpass the helium regime of low temperatures. As per the phase diagram of hydrogen, 15 K could be achieved using liquid hydrogen. In the 1950s, Matthias and Hulm began systematic search for new superconductors and delivered over 3000 different superconducting alloys. Matthias was recognized as a wizard in the field of superconducting materials. With his experience, he provided 5 rules to discover new superconductors: take material with high crystal symmetry, material with high density of states at Fermi level, avoid oxygen, avoid magnetism and  no insulators. Later on it was realized that deviation from such constraints could in fact lead to potentially more interesting superconductors. Despite of his unofficial (and somewhat provocative) sixth rule ``Stay away from theorists", in 1957 a microscopic theory of superconductivity was proposed by Bardeen, Cooper and Schrieffer. This came to be popularly known as the BCS theory. The BCS theory explained the origin of superconductivity in simple systems like elemental metals, binary alloys etc. However, with the discovery of high temperature cuprate (and, later the pnictide) superconductors, physicists soon realised that superconductivity in general cannot be explained in the framework of BCS theory. Thus theoretical research spurred to identify a common mechanism that would explain superconductivity in low $T_c$ as well as high $T_c$ materials. The search for new theories also provided innovative ideas for exciting experiments, development and integration of various concepts and that led to the discovery of a variety of different superconducting materials. The most recent additions in the list of exotic superconductors, perhaps, are the so-called topological superconductors. 

In general, the superconducting ground state is described by the Bogoliubov-DeGennes Hamiltonian. The positive and negative energy eigenstates of the Bogoliubov-DeGennes Hamiltonian in superconductors appear in pairs. When the superconducting condensate is formed, the negative-energy eigenstates remain fully occupied. This enables one to define various topological numbers (e.g., the Chern number) for the occupied states, as it is done in insulators with non-trivial band structure topology. Non-zero topological numbers can exist along with a fully formed superconducting energy gap in the bulk of the superconductor. Due to topological restrictions, gap-less ``Majorana" modes may also appear in the surface of such superconductors, thereby making the surface effectively ``gap-less". Such a superconductor is categorized as a topological superconductor.
\\
It has been widely believed that one way of achieving achieving topological superconductivity could be simply doping charge carriers through metal intercalation in materials with topologically non-trivial band structure. Significant success in terms of making topological materials through controlled doping of topological materials could be achieved. For example, topological insulators like Bi$_2$Se$_3$ became superconducting upon intercalation with ions of certain metals like Cu, Sr and Nb.  It has also been found that at the required doping level for superconductivity ($\sim$ 2 $\times$ 10$^{20}$ cm$^{−3}$ ) in such doped topological systems, the topological surface states are still well separated in the momentum space from the bulk states. Therefore, it is expected that the surface may also become superconducting due to the proximity of the bulk superconductor thereby giving rise to a 2D topological superconducting phase on the surface. However, experimental confirmation of the same remained an unattained goal. Most of the spectroscopic measurements indicated that such induced superconducting phases appear to be conventional BCS like. In case of Nb-Bi$_2$Se$_3$ deviation from BCS-like behavior was found, but no clear signature of Majorana zero modes were found.

Another promising way of inducing superconductivity in a topological material could be through applying pressure. A pressure-induced superconducting phase was found in Bi$_2$Se$_3$ even in absence of any doping. However, this superconducting phase was also found to be BCS-like. In Sr-Bi$_2$Se$_3$, an interesting pressure-induced superconducting phase was found. The ambient superconductivity of this system was first seen to disappear with applying pressure and emerge again at higher pressures. The high-pressure re-entrant superconducting phase was found show a significantly higher $T_c$ compared to the $T_c$ of the ambient superconducting phase. The re-entrant  phase showed strong signatures of unconventional superconductivity, but again, from the experimental data, it was not possible to conclude whether the superconducting phase would satisfy the criteria of a topological superconductor.

In this review, we will discuss about another novel way of inducing superconductivity in a topological material. In 2014, while performing point contact spectroscopy experiments on Cd$_3$As$_2$, the researchers at IISER Mohali, India found that mesoscopic point contacts between Cd$_3$As$_2$ and sharp tips of metallic silver (Ag) showed superconductivity-like features. By performing point contact spectroscopy in various regimes of mesoscopic transport, the researchers successfully showed that the phase was indeed superconducting\cite{arxiv1}.This was later published in Nature Materials\cite{Aggarwal2}. But, such a superconducting phase appeared only at mesoscopic junctions made with normal metals on Cd$_3$As$_2$. Cd$_3$As$_2$ is known to be a topological Dirac semimetal. Later on, such a mesoscopic superconductng phase was discovered in a wide variety of topological semimetals. The phase came to be popularly known as tip-induced superconductivity (TISC).

Since the TISC phase emerges only at a confined region under a mesoscopic point contact, traditional bulk tools for characterizing  superconductivity cannot be employed. In such a confined geometry neither a zero resistance state can be directly measured, nor sufficient magnetic signal can be obtained to detect the existence of perfect diamagnetism. However, such a point contact geometry is ideal for probing the other standard hallmark signatures of superconductivity, like the existence of a temperature and magnetic field dependent superconducting energy gap and a temperature and magnetic field dependent critical current. We will review the details of the experimental signatures that can be used to prove the existence of superconductivity even when the text-book tests cannot be performed. Then, we will review different systems where a TISC phase could be realized. We will also discuss the knowledge that is gained on the nature of superconductivity in TISC realized on various systems.

 \section{Point contact spectroscopy}
 
Before discussing TISC, it is important to give an overview a special transport spectroscopic technique called point contact spectroscopy. This discussion will help us understand the logical sequence that is exploited to establish the existence of TISC unambiguously.

\subsection{Fabrication of point contacts}
A point contact between two different materials can be formed in a number of ways. Perhaps, the most common method of making a point contact involves slowly moving a sharp metallic tip towards the surface of a sample until a light ``ohmic" contact is established. This method is popularly known as the ``needle-anvil" \cite{Naidyuk,Daghero,Janson} method. Here, the tip can be moved towards or away from the sample surface using either a mechanical differential screw arrangement or piezo driven mechanism(See Fig1(a)). Both of these  mechanisms provide great control over the cross-section of the micro-constrictions formed under the point contacts. Another way of forming a point contact involves pasting a thin(~25 $\mu$M) Gold wire on sample surface using a small(~50 $\mu$M) Indium flake or drop of Silver paste. The contact diameter here can be tuned by application of current or voltage pulses across the contact region. The application of a voltage pulse causes, either creation of new conduction channels or annihilation of existing ones, via piercing of the oxide layer present on the surface of either of the materials forming the contact, there by increasing or decreasing the cross-sectional area of the constriction. This method is commonly known as the ``Soft contact" method since these contacts do not impart any pressure on the sample\cite{Naidyuk,Daghero}. Soft contacts are more stable both thermally and mechanically than contacts formed by needle-anvil method. Point contacts can also be made by breaking or shearing \cite{Naidyuk,Daghero} a material so that the two pieces of a given material are barely connected forming a micro-constriction. However this method is highly destructive and there is minimal control over the contact cross-section. This kind of micro-constriction can also be fabricated by using lithographic techniques. However for this type, each fabricated micro-constriction has a fixed cross-sectional area. Nevertheless, the need-anvil method is non-destructive and provides huge control over the cross-section of the micro-constriction, hence is most widely used.

\begin{figure}[h]
\centering
\includegraphics[scale=0.8]{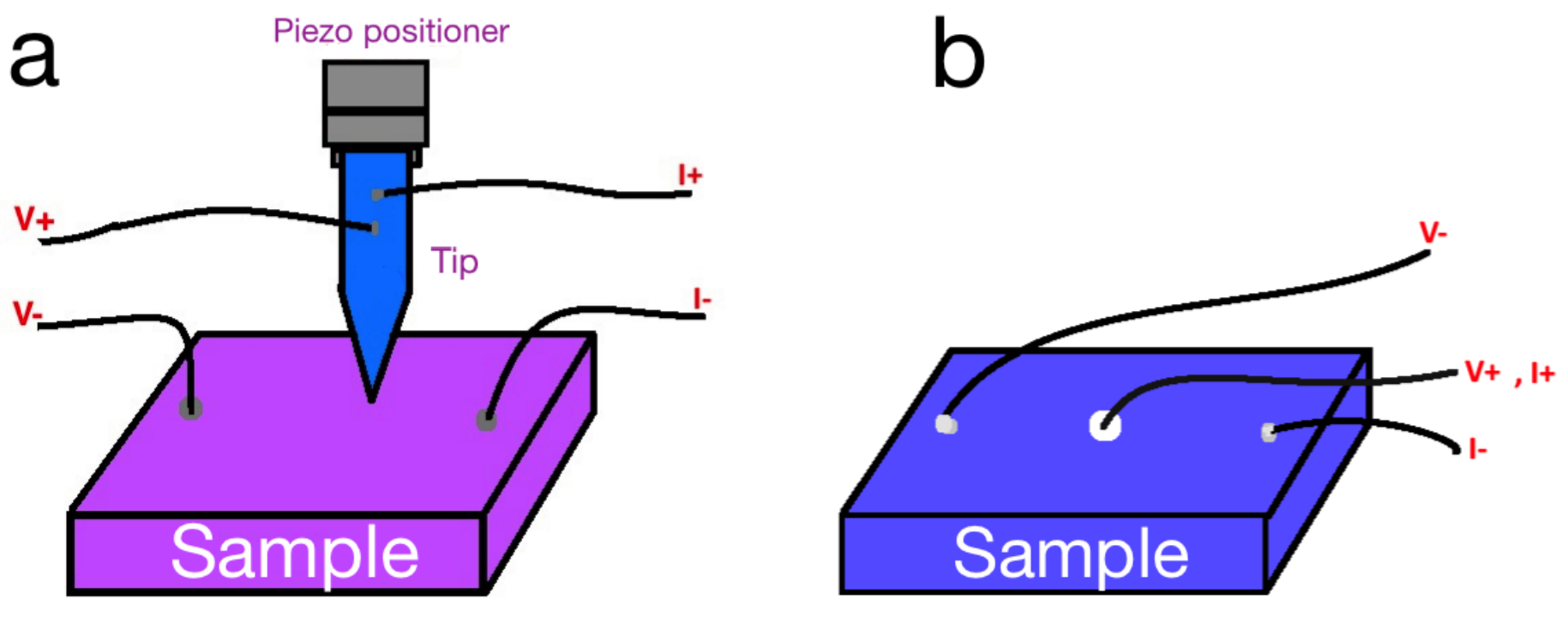}
\caption{Schematic of a point-contact set up with electrical connections. \textbf{(a)} The needle-anvil method: the point-contact is made by approaching a sharp tip (needle) towards the sample by manual differential screw or piezo-controlled nano-positioner. (\textbf{b)} The ``soft'' point-contact: Using small In-flake or drop of Ag-paste is used to make a soft point-contact on the sample.}
\label{pc}
\end{figure}

\subsection{Electrical measurements}


\begin{figure}[h]
\includegraphics[scale=0.8]{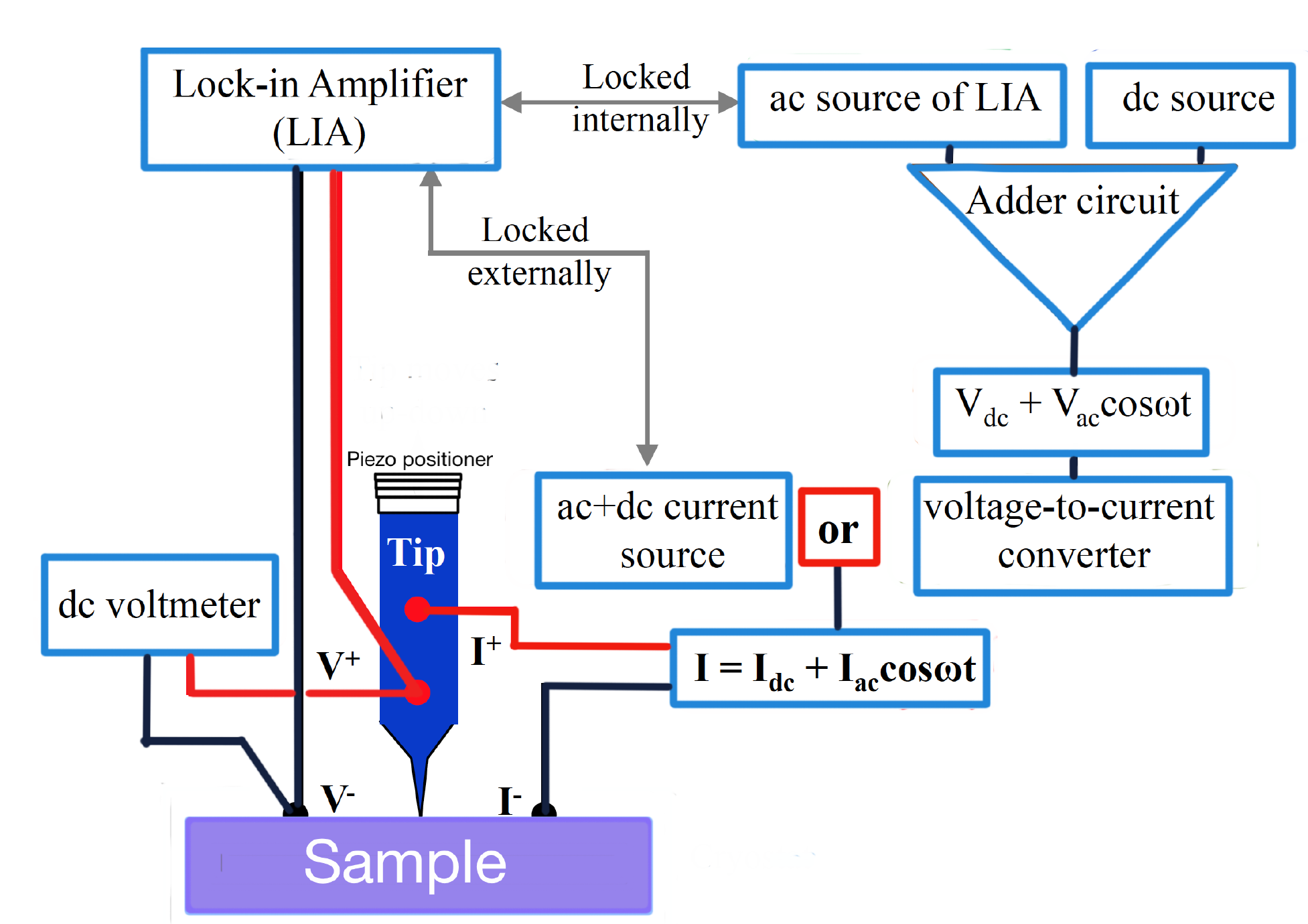}
\caption{Schematic of PCS measurement using lock-in based modulation technique.}
\label{crkt}
\end{figure}

A point contact spectroscopy experiment involves formation of a micro-constriction between two materials and then analyzing the $I-V$ characteristics across the constriction to extract information about the Fermi surface properties of the materials. Ultimately, the non-linearities, if any, in the $I-V$ characteristics provide the desired spectroscopic information. The signature of such non-linearities can be seen in the derivatives of a $I-V$ curve, i.e., a $dI/dV$ vs. $V$ spectrum. Such a spectrum can be obtained by digitally differentiating a $I-V$ curve. However, in doing so, the calculated spectra can be noisy. The most common technique for directly obtaining $dI/dV$ vs. $V$ spectrum for a point contact is a lock-in amplifier based modulation technique(See Fig2 ). In this technique, a small dc current ($I_{dc}$) is coupled with a small amplitude, low  frequency ac current ($I_{ac} cos\omega t$) ($I_{dc} >> I_{ac}$) and is then passed through the point contact. The corresponding  dc potential drop across the contact is simultaneously recorded. The dc current is supplied through a dc current source while the ac current is provided through a Lock-in amplifier system by utilizing a voltage to current converter. The dc current is swept in small steps while the ac current is kept fixed (i.e. the frequency and amplitude is kept constant). Usually the frequency of applied ac current is in the range of a few  Hz to a few kHz. Application of frequency in multiples of line frequency must be avoided and use of noise filters is recommended to minimize noise level.  The dc potential drop across the point contact is recorded using a voltmeter (or a digital multimeter) and the differential conductance/resistance is determined from the ac output voltage that is measured by the lock-in amplifier. The lock-in amplifier signal locked at the first harmonic is proportional to $dV/dI$ and the second harmonic is proportional to (d$^2$V/d$I^2$) and so on. This can be easily seen by expressing the output voltage as Taylor series expansion.\\
 $V(I=I_{dc}+I_{ac}cos\omega t)=V(I_{dc})+\left(\frac{dV}{dI}\right)\mid_{I_{dc}}I_{ac}cos\omega t+\frac{1}{2}\left(\frac{d^2V}{dI^2}\right)\mid_{I_{dc}}(I_{ac}cos\omega t)^2+\cdots=V(I_{dc})+\left(\frac{dV}{dI}\right)\mid_{I_{dc}}I_{ac}cos\omega t+\frac{1}{4}\left(\frac{d^2V}{dI^2}\right)\mid_{I_{dc}}(I_{ac})^2(1+cos2\omega t)+\cdots$ \\

\subsection{Different regimes of point-contact transport}

In general, electronic transport through a point contact depends on the type of the materials forming the point contact. Since the charge carriers need to be transported through narrow constrictions, the dimensions of the constriction also plays a prominent role in such transport. Depending on how large or small a point contact is in comparison to the characteristic length scales of the charge carriers (e.g., the electronic mean free path $l$) transport through a point contact can happen in different mesoscopic regimes as discussed below.\\
        
        \textbf{a. The extreme quantum regime:} When the diameter of a point contact ($a$) is comparable to or of the order of few de Broglie wavelengths of the electron, the electronic transport happens in the so-called ``quantum regime".  Here, the electrons conduct through quantized conduction channels each of which contributes one conductance quantum $G_0 = \dfrac{2e^2}{h}$ to the total point contact conductance. To understand this in simple terms, one may use simple example of an hourglass (an hourglass consists of two pear shaped glass bulbs connected to each other at the apex with a minute passage between them ). Consider a hourglass where each sand grain is identical(spheres of same radius). Suppose the passage at the apex has the same diameter as of the sand grains then, only one sand grain can pass through the passage at a time. This is same as  electrons passing through a single conduction channel and contributes single conductance quantum, since only one electron is passing through a section at a time. Now if the diameter of passage is doubled then at a time, at most, only two sand grains can pass through( and conductance quantum number is 2). Similarly if the diameter is tripled then at an instant at the most 3 sand grains can pass( quantum number =3) and so on. This mimics the flow of electrons through quantized conduction channels. It can be easily seen that if the diameter of the passage was sufficiently large then all of the sand grains could pass at the same time and reducing the diameter only served to impose resistance to flow of sand through the passage. With the maximal resistance is imposed when there is only one conduction channel. Though this argument sounds classical in nature, this helps understand the origin of Sharvin's resistance for electron conduction through narrow constrictions.  The conductance is equal to $N.G_0$ for N conductance channels. The conductance quantization is observed as conductance steps in conductance vs. contact diameter plots. \\
        
        \textbf{b. The ballistic regime:} The transport of electrons through individual conduction channels is ballistic in nature. That means, the electrons are transported without any scattering. In general, how far the ballistic transport is carried out depends on one of the characteristic length scales, the elastic mean free path of charge carriers and the dimensions of the medium through which the charge carriers travel.  For metallic point contacts ballistic transport is favored whenever the contact diameter $a$ is much smaller than elastic mean free path of the electrons($\lambda_{de-Broglie}<<a<<l_e$). Due to large mean free path and smaller contact diameter the electrons statistically do not undergo any scattering inside the contact region and  travel ballistically. In other words, the collisions of the electrons occur far away from the contact region causing a dissipation-less current to flow through  the contact. Absence of dissipation in a ballistic point contact makes point contact spectroscopy a powerful spectroscopic probe. A dc bias across a point contact in the ballistic regime is used by the electron accelerate to a high kinetic energy within the contact region. At such high energies it can then excite elementary excitations like the phonons in a metal, magnons in a magnet or more complex entities like the Bogoliubuons in superconductors. Such excitations, in turn, cause the electron to get scattered thereby giving rise to non-linearities in the $I-V$ characteristics. That is the basis of energy resolved point contact spectroscopy. With single crystalline samples, point contact spectroscopy can also be used as a momentum-resolved spectroscopic probe. With superconducting point contacts, spin-resolved information can also be extracted. We will discuss this in more detail later. In the ballistic regime, the point contact resistance is purely given by Sharvin's resistance formula\cite{Sdutta}.  $Rs =\dfrac {2h}{e^2 (ak_f)^2}$, where $k_f$ is the Fermi momentum. To note, the bulk resistivity of the materials forming the point contact do not appear in the formula. \\

        \textbf{c. The diffusive regime:}  In this regime, the contact diameter is larger than $l_e$ but is smaller than the diffusion length $\lambda_D= (l_el_{in})^{1/2}$. Here, elastic scattering of the electron with other objects is allowed. Hence, it is possible to only obtain energy resolved spectroscopic information in this regime of point contact transport -- momentum information is completely lost.

         \textbf{d. The Thermal regime:} For electrons travelling through a contact with contact diameter greater than the inelastic mean free path of electrons ($a<< l_e,l_{in}$), one enters the thermal regime of transport.  In this regime, the electrons feel a drift due to applied voltage bias and move through the contact however, the motion is disordered and inelastic collisions occur within the contact region. This results in dissipation of energy in form of heat and leads to heating of the contact. The temperature at the centre of the contact is maximum and is given by $T^2_{max}=T^2_{bath}+\dfrac {V^2}{4L}$, where $T_{bath}$ is the bath temperature and $L$ is the Lorenz number. Due to heating of contact the energy resolved information is lost and makes it unsuitable for performing energy resolved spectroscopy. In this case, Maxwell considered the point contact as a circular opening (orifice) of diameter 2$a$ through which a current flows as a potential is applied across the contact. By taking appropriate approximations he showed that the maximum contribution to the point contact (Maxwell's) resistance $R_M=\rho/2a$ comes from the centre of the point contact, where $\rho$ is the resistivity of the material. This resistance is usually known as the spreading resistance and is associated with the spread of current as it moves away from the orifice. By measuring the Maxwell's resistance ($R_M (V)$ one actually measures the temperature dependent resistivity of the material. \\
         
         \textbf{e. The intermediate regime:} Depending on the mean free path and contact diameter, it is possible to enter an intermediate regime of transport where the contact diameter is such that both Sharvin's resistance and Maxwell's resistance components of the total resistance become comparable. The total resistance of the contact is given by the Wexler's formula\cite{Wexler} $R_W= \dfrac {2h}{e^2 (ak_f)^2} +\Gamma(\dfrac{l}{a}) \dfrac {\rho}{2a} $,
        $\Gamma$ is a slowly varying function of the order of unity. 

\section{Point contact Andreev reflection (PCAR)}

For point contacts between a normal metal and a superconductor, the electronic transport in the ballistic regime is dominated by a quantum process called Andreev reflection. In order to understand the process, let us consider the density of states of a conventional superconductor with energy in a so-called semiconductor model. In fig. 3 we depict an illustration of basic features of BTK model where, on the right hand side we have shown the density of states(DOS)profile of  a conventional superconductor. The DOS for the superconductor diverges at E =eV=$\pm \Delta$ at bias voltage V. This result can be derived from the expression for quasiparticle DOS working under the axioms of BCS theory where the DOS is given by N(E) = Re$\dfrac{E}{\sqrt{E^2-\Delta^2}}$. Typically, for conventional superconductors the value of energy gap ($\Delta$) is found to be in the range of a few meV.   The DOS in metallic side in this energy range can be approximated as a constant and is represented by a flat line  as seen in the left hand side of the fig.3. 
 Now suppose an electron approaches towards the N/S interface from the metallic side with an energy E$_1$. This electron finds available states on the superconducting side and can transports normally through the interface. Now suppose an electron with energy E$_2$ approaches the interface. This electron due to having an energy which lies in the energy gap in superconducting side, cannot find available states and cannot transmit through the interface. For such a scenario, the electron can cross-over through Andreev reflection process. The electron travelling through the interface now, must pair up with another electron of opposite spin to form a cooper pair and travel inside the superconductor. During this process, to conserve the angular momentum a hole is reflected back into the normal metal. The reflected hole is termed as an Andreev reflected hole. Through the Andreev reflection process a normal current flowing through the N/S interface gets converted into a super-current. 

.

A conductance mapping of the electronic transport through a mesoscopic N/S interface i.e. a superconducting point contact, with respect to the applied voltage bias, yield certain characteristic spectra where the consequences of Andreev reflection are visible. Such spectra can be analysed under a suitable model to extract energy and momentum resolved information about various scattering mechanisms occuring inside a mesoscopic region around the interface. Hence,  coupled with our knowledge of electronic transport through a micro-constriction, information obtained from Point contact Andreev reflection can be used as an effective spectroscopic tool to investigate the electronic transport through a mesoscopic point contact between superconductor and metals.\\
The contribution of AR in electronic transport can be most prominently observed in superconducting point contacts where electron transport is ballistic in nature. For such contacts, contribution due to AR manifests as two sharp peaks, symmetric about the zero voltage bias at  $+\dfrac{\Delta}{e}$ and $-\dfrac{\Delta}{e}$

\subsection{Andreev reflection as a spectroscopic tool}
   
\subsubsection{\textbf{Blonder-Tinkham-Klapwijk theory}}

In the ballistic regime of transport for a N/S point contact, the differential conductance(AR) spectra is analyzed under the BTK model. The BTK model assumes the potential barrier as a delta function of the form H= V$_o$ $ \delta(x)$. In fig.3, the interface is placed at x=0 .The barrier strength is characterized by a dimensionless parameter,Z=$\dfrac{V_o}{\hbar v_f}$, for mathematical convenience. This parameter originates primarily due to two reasons: a) Fermi velocity mismatch between the materials forming the contact which suggests that the barrier can never be transparent for conduction. and (b) due to presence of naturally occurring oxide layer on the surface of materials, the to materials forming a point contact will have an effective layer of oxide material acting as a barrier. Hence, the o presence of a potential barrier presents a finite probability for the electrons to undergo normal reflection from the barrier along with Andreev reflection.   The expression for current through the point contact now requires inclusion of outgoing and incoming populations. The Andreev reflection and normal reflection probabilities are represented by A(E) and B(E) respectively in the expression for current.\\
 \begin{figure}[htb]
	\centering
\includegraphics[scale=0.6]{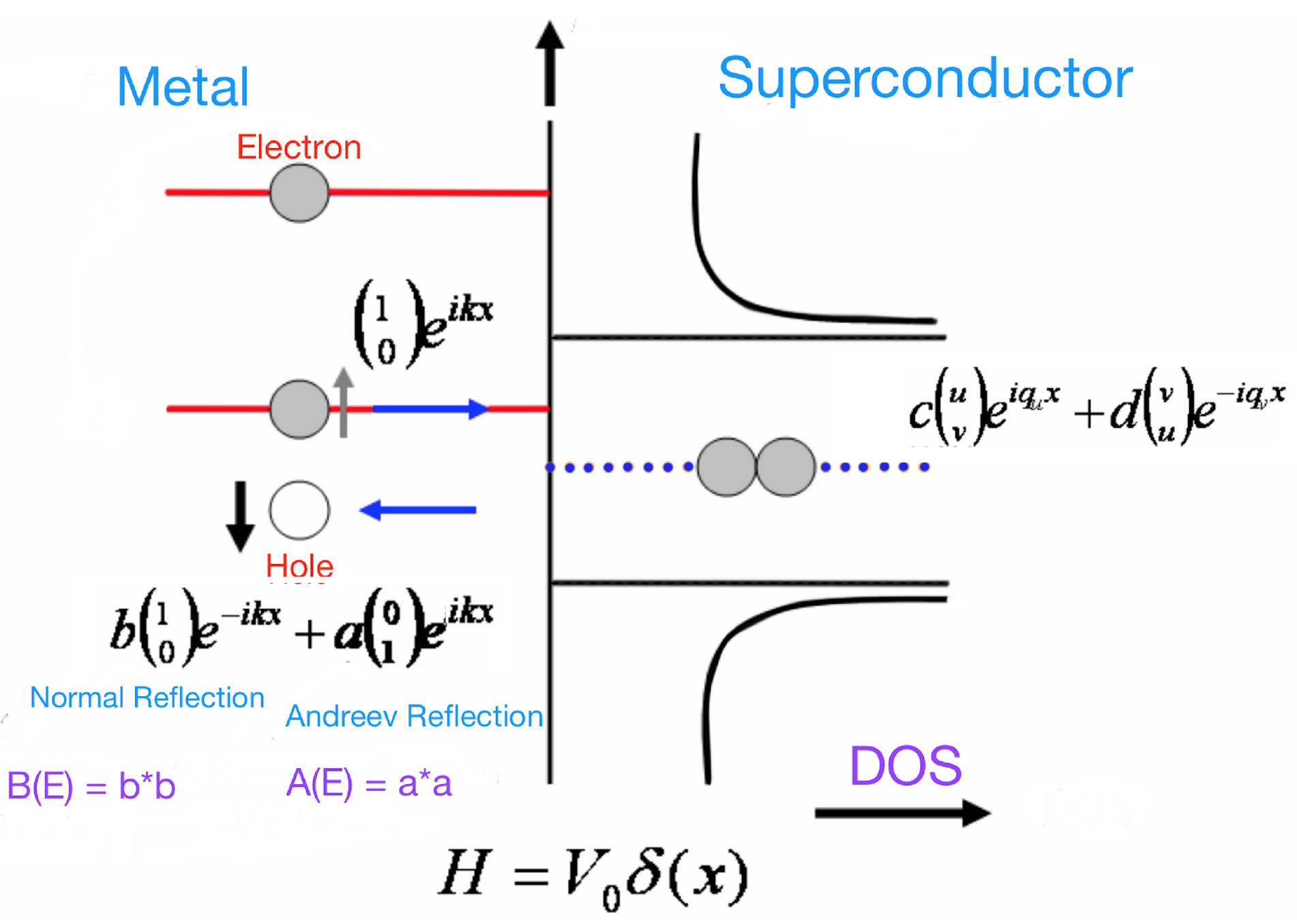}
	\caption{BTK theory- Andreev reflection process represented via dos profile for a normal metal- superconductor interface}
	\label{pb1}
\end{figure}\\

 $I_{NS} \propto N(0). v_F \int_{-\infty}^{+\infty} [f_0(E-eV)-f_0(E)][1+A(E)-B(E)] dE$\\
 where, N(0) is the density of states at the Fermi level and $v_F$ is the Fermi velocity.  A(E) and B(E) can be easily calculated by matching the boundary conditions for a delta function potential with wave functions that satisfy the Bogoliubov-deGennes (BdG) equations.\\
 
 $i\hbar \dfrac{\partial f}{\partial t}=[-\dfrac{\hbar^2 \nabla^2}{2m}-\mu(x)+V(x)]f(x,t)+\Delta(x)g(x,t)$,  and \\
 
 $i\hbar \dfrac{\partial g}{\partial t}=-[-\dfrac{\hbar^2 \nabla^2}{2m}-\mu(x)+V(x)]g(x,t)+\Delta(x)f(x,t)$ \\
 
where, $\Delta$(x) represents the superconducting energy gap, $\mu$ is the chemical potential, f(x,t) and g(x,t) are the elements of column vector $\psi$ describing one type of BCS quasiparticle states. 
For constant $\mu$(x), $\Delta$(x) and V(x) we can use the following trial solutions \\
$f(x,t) = ue^{(ikx-iEt)/ \hbar}$ and $g(x,t)=ve{(ikx-iEt)/ \hbar}$\\

When V(x) =0, E=$\sqrt{\left[ \dfrac{\hbar^2k^2}{2m}-\mu \right]^2}$
For energies greater than 0, we have

$u^2=\dfrac{1}{2} \left[ 1 \pm \dfrac{(E^2-\Delta^2)^{1/2}}{E}\right]=1-v^2$\\
Now if the wave functions inside the normal metal and superconductor is represented by $\psi_N $ and $\psi_S$ respectively, it is possible to obtain the Andreev reflection A(E) and normal reflection B(E) probabilities(provided in table below).

\begin{table}[ht]
     \centering
     \begin{tabular}{|c|c|c|}
     \hline
     Coefficient  &  $ E<\Delta$ &  $E>\Delta$\\
     \hline
    $ A(E)$   &  $\dfrac{(\Delta/E)^2}{1- \epsilon(1+2Z^2)^2}$ &   $\dfrac{(uv)^2}{\gamma^2}$\\
    \hline
    $B(E)$ & $1-A(E)$ & $\dfrac{[u^2-v^2]^2Z^2(1+Z^2)}{\gamma^2}$\\
    \hline
     
   \end{tabular}  
   \caption{Table showing Andreev and normal reflection probabilities for an unpolarized system.}
\end{table}

\subsubsection{\textbf{spin polarized BTK theory}}
Recently, Point contact Andreev reflection(PCAR) technique has proved to be one of the widely used, efficient and reliable spectroscopic tool with a tremendous capability to investigate  electronic properties of a variety of materials. In the previous section we have described the BTK formalism, utilized to analyze the PCAR spectrum,  to extract energy and momentum resolved information. In practice, for a N/S point contact, the PCAR spectrum is fitted with a simulated curve obtained by utilizing the BTK formalism where, one can use the superconducting energy gap($\Delta$), the potential barrier strength(Z) and a broadening parameter ($\Gamma$) within some predefined constraints. PCAR spectroscopy can also be used to investigate magnetic materials to ascertain the degree of spin polarization. This gives us a new technique known as spin resolved PCAR. \\

. \begin{figure}[htb]
	\centering
	\includegraphics[scale=0.8]{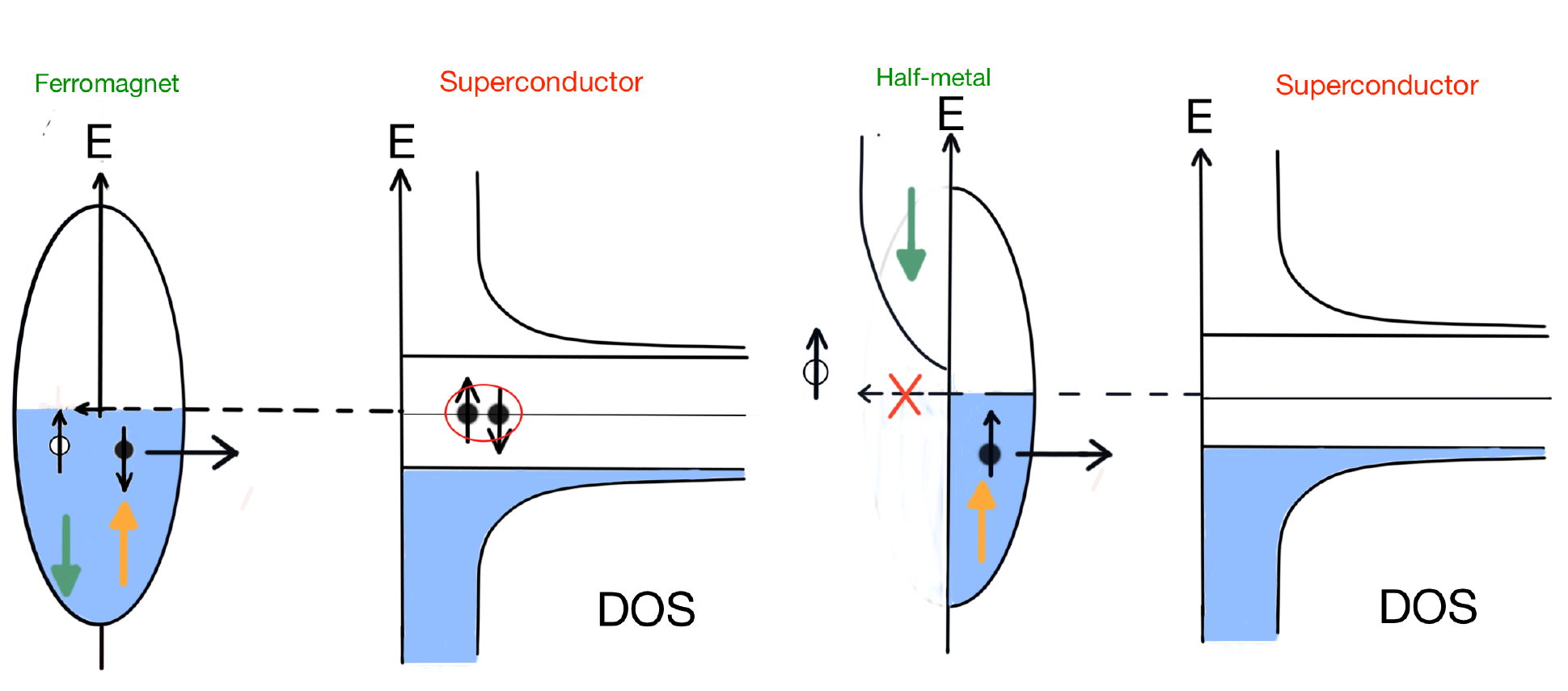}
	\caption{spin polarized BTK theory- Andreev reflection process represented via dos profile for a ferromagnet- superconductor interface}
	\label{pb1}
\end{figure}

  Spin resolved PCAR is one of the most favored techniques to measure the transport spin polarization of materials. This method is independent of sample geometry, has excellent energy resolution and does not require application of magnetic field. One may consider the possibility of surface modification, due to naturally occurring oxides at surfaces or due to chemical reactions at surfaces of superconductor and ferromagnet, as a disadvantage of the technique. However, since point contact is established after the tip penetrates the surface oxide layer, the effect it is probable that the effect of surface modification is negated and the results obtained arise purely from formation of contact between the intended material. A confirmation to this is provided by the good agreement of results obtained via other methods. 
The BTK formalism can be extended to analyze PCAR spectrum obtained for a ferromagnet/superconductor(FS) point contact where the presence of spin-polarized current necessitates the use of additional fitting parameter, namely the spin polarization (P$_C$). Now let us take a closer look at what goes on at a ferromagnet/superconductor interface in terms of density of states. The Fermi level of a ferromagnet/superconductor point contact is expected to be spin polarized i.e at the Fermi level the DOS of down-spin(N$_\downarrow$) and up-spin(N$_\uparrow$) electrons are not equal. This implies that $|N_\uparrow$-N$_\downarrow|$ electrons approaching the interface cannot undergo Andreev reflection due to unavailability of accessible states in the opposite spin band(Fig.4). This causes suppression of Andreev reflection in ferromagnet/superconductor point contacts. The BTK formalism in this case relies on determining this suppression in Andreev reflection to extract the degree of spin polarization of the Fermi surface. The Andreev reflection and normal reflection probabilities for this case in listed out in Table II.\\
The steps followed to extract the absolution value of spin polarization of Fermi level are: 1) start with calculation of BTK current for  zero spin polarization (I$_{BTKu}$ and for a fully(100\%) spin polarized case(I${BTKp}$). 2) Then calculate the current for an intermediate spin polarized(P$_t$) case by interpolating the current value between I$_{BTKu}$ and I$_{BTKp}$ by following the relation I$_{total}$= I$_{BTKu}$(1-P$_t$)+ P$_t$I$_{BTKp}$. 3)one just needs to take a derivative of I$_{total}$ with respect to applied bias V to get the modified Andreev reflection spectrum having signatures of spin polarization in the ferromagnet. Once again we emphasize that for fitting of such a spectrum under the modified BTK formalism\cite{Aggarwal1, Mazin, Soulen} requires use of four fitting parameters Z, $\Delta$, $\Gamma$ and P$_t$.

\begin{table}[ht]
     \centering
     \begin{tabular}{|c|c|c|}
     \hline
     Coefficient  &  $ E<\Delta$ &  $E>\Delta$\\
     \hline
    $ A(E)$   &  $0$ &   $0$\\
    \hline
    $B(E)$ & $1$ & $\dfrac{(\epsilon^{1/2}-1)^2+ 4Z^2\epsilon}{(\epsilon^{1/2}+1)^2+ 4Z^2}$\\
    \hline
     
   \end{tabular}  
   \caption{Table showing the values of Andreev reflection and normal reflection probabilities for a polarized system. }
\end{table}

\subsubsection{\textbf{Broadening of DOS}}

The PCAR spectra obtained for point contact between normal metals and conventional superconductors are usually in good agreement with the BTK theory. However, the same cannot be said for certain point-contacts between a) normal metals and unconventional superconductors and, b) superconductors and ferromagnets where the experimentally obtained spectra is found to be broadened from the theoretical simulation. It is believed that this broadening occurs due to shortening of quasiparticle lifetime resulting due to inelastic scattering of charge carriers near the contact region. This warrants incorporation of this effect into the BTK formalism. The broadening is taken care of by including the inelastic term $\Gamma=\dfrac{\hbar}{\tau}$ in the Boguliobov de Gennes equations where, $\tau$ is the quasiparticle lifetime.  The inclusion of the broadening parameter modifies the BdG equations which are\\

$i\hbar \dfrac{\partial f(x,t)}{\partial t}=\bigg[\dfrac{\hbar^2 \nabla^2}{m}+\mu+i\Gamma -V(x)\bigg]f(x,t)+ \Delta g(x,t)$
 \\
 \\
 $i\hbar \dfrac{\partial g(x,t)}{\partial t}=\bigg[\dfrac{\hbar^2 \nabla^2}{m}+ \mu+i\Gamma-V(x)\bigg]g(x,t)+\Delta f(x,t)$
\\
\\
The trial wavefunction now contains a decaying term e$^{-\Gamma t/ \hbar}$ and the pair occupation probabilities get modified to 
$u^2=\dfrac{1}{2}\bigg[1\pm\dfrac{((E+i\Gamma)^2-\Delta^2)^{1/2}}{E+i\Gamma}\bigg]=1-v^2$

resulting in modification of Andreev and Normal reflection probabilities.
\\
$A(E)=\dfrac{\sqrt{(\alpha^2+\eta^2)(\beta^2+\eta^2)}}{\gamma^2}$ and \\
$B(E)=Z^2\dfrac{[(\alpha-\beta)Z-2\eta]^2+[2\eta Z+(\alpha-\beta)]^2}{\gamma^2}$
\\
\\
Hence, $\Gamma$ is incorporated in the BTK model as an additional fitting parameter to take care of broadening in the Andreev reflection spectrum. (
 The modified DOS at energy E is given by $N(E) \sim Re \bigg(\dfrac{E+i\Gamma}{\sqrt{(E+i\Gamma)^2-\Delta^2}}\bigg)$. 

\subsection{The role of critical current}

Point contact spectroscopy exploits the characteristic properties of ballistic point contact between two materials to obtain energy, momentum and spin resolved spectroscopic information. In the previous sections we have discussed that the  Andreev reflection spectrum is analyzed under the BTK model where it is assumed that the two materials( a superconductor and normal metal) form a ballistic interface. Earlier we have seen that a conventional PCAR spectrum exhibits only two peaks at $\pm \dfrac{\Delta}{e}$ respectively and beyond this voltage only smooth decay of DOS is observed. The maximum zero bias enhancement in differential conductance of a factor of 2 is possible only at zero temperature and for a completely transparent barrier (Z). However, in practice certain additional features are observed in the conductance spectrum such as multiple dip and peak structures, anomalous large zero bias enhancement etc. Previously these additional features were attributed to exotic physical phenomena such as proximity induced, energy gap signature, inter-grain Josephson effect, unconventional pairing, presence of multiple order parameters, Andreev bound states etc. These explanations were system specific but the observation of these structures in wide variety of materials indicated towards the possibility of a more general explanation. A series of PCAR measurements performed by sheet et. al. on several combinations between metal and elemental superconductors where these dip structures in differential conductance spectrum were repeatedly observed. Moreover, these dip structures showed systematic dependence on temperature and applied magnetic field. Analysis of these structures on wide number of point contacts helped the author to scrutinize the previous explanations. For example, absence of these dips in contacts between a ferromagnet and superconductor was earlier thought to be confirmation of proximity induced superconductivity resulting in dip structures, however, sheet et. al. observed that the dip structures appeared consistently for point contacts between superconductors and other non magnetic metals. Similarly, the hypothesis that the dip structures appear due to inter-grain Josephson junction was ruled out due to repeated observation of such structures in single crystalline superconductors. These dip structures are not bias specific and were observed to appear at various bias values though always symmetric about the zero bias. The systematic evolution of dip structures with temperature and magnetic field confirmed that these structures are unambiguously related to superconductivity. The dips were also observed to evolve with the contact diameter. The dips were observed to diminish with increasing contact resistance i.e for decrease in point contact size for various combinations of metals, superconductors and ferromagnets forming point contact. These experiments revealed a clear correlation between dip structures and contact resistance or contact size. These results ultimately led to the development of a theoretical model that includes the role of critical current. \\
The model distinguishes between different type of spectra that are commonly observed in superconducting point contacts. The distinction is made on the basis of contact diameter and the electronic mean free path. This has been discussed previously in section titled " Different regimes of point-contact transport". We briefly revisit the salient features of this model here. The transport through a point contact is classified primarily three different regimes on the basis of contact diameter and mean free path of electrons. On one extreme we have the ballistic limit, where the contact diameter is much smaller than the elastic mean free path so that, statistically speaking,  there is absence of scattering within the point contact regime. This allows the electrons to travel ballistically within the contact. Therefore only structures correponding to AR namely the Andreev peaks appear in the differential conductance spectrum. In this case the resistance in contact arise purely due to Sharvin's contribution to resistance.  On the other extreme we have the thermal limit, where the contact diameter is much larger than the inelastic mean free path. Hence, in this limit all scattering mechanism take place within the contact region resulting in dissipation of energy in form of heat. In this limit the bulk resistivity of metal plays key role in deciding the contact resistance  and is given by Maxwell's resistance formula. In this limit, the differential conductance spectra exhibit two dip structures symmetric about the zero voltage bias. If the contact diameter is larger than the elastic mean free path but smaller than inelastic mean free path, then one reach the intermediate state where   both of the structures discussed above appear simultaneously in the differential conductance spectrum. The contact resistance in this case has contribution from both the Sharvin's and Maxwell's resistance. A look at the I-V characteristics shows a sharp change in slope at bias corresponding to the critical current of the superconductor. The dips in differential conductance spectrum appear at the exact same value of bias corresponding to the sharp change in slope in I-V. Moreover, when the current corresponding to the critical current of the superconductor flows through the contact, the resistance of superconductor increases sharply to attain the normal state value resulting in huge zero bias enhancement in normalized differential conductance. Hence it is clear from these above discussion that the  dip structures in differential conductance spectrum correspond to the critical current of the superconducting point contact. In this analysis of point contact spectra, the effect of contact heating on critical current is neglected as it arise due to contribution from Maxwell resistance. This assumption works well for contacts where R$_M$/R$_S$ is small. For contacts where R$_M$/R$_S>>1$, a significant heat dissipation occurs at contact and the effective temperature rises rapidly with applied bias. The heat dissipation at low temperatures may drive the superconductor into normal state at current values lower than the critical current of superconductor. \\
It is possible in principle to think of a situation where for a contact in ballistic limit, the critical current is reached at smaller voltage values ($\sim$ $\Delta$/e). For this case, feature associated with Andreev reflection is not observed since they are supposed to occur in voltage range $\pm 2\Delta/e$ and one gets a spectrum with only sharp dips. This possibility might be seen in superconductors having low critical current densities or while working at temperatures close to T$_C$ where critical current is small. \\
\\\\
 So far, we have looked at the hallmark signatures of superconductivity that appear in transport spectroscopy of a superconducting material in a point contact geometry. We have classified the different types of spectra that are obtained for such a contact into broadly into three categories on basis of contact diameter and mean free length namely 1) Ballistic regime, 2) Intermediate regime and 3) Thermal regime. We have also understood the origin of the spectroscopic features appearing in spectra obtained in these regimes. In the following sections we describe how to characterize a point contact exhibiting TISC.


\section{Tip-induced superconductivity (TISC)}

With the understanding of the characteristics features that are obtained in the different transport regimes of point contact spectroscopy involving superconductors, it was possible to detect tiny volumes of superconducting phases that appeared only under the point contacts made with a normal metal on certain complex quantum materials, specially those with definite topological features. Such phase was named as tip-induced superconductivity (TISC) by the Indian group that discovered the TISC phase on Cd$_3$As$_2$ for the first time. Following that, a lrge number of groups around the world realized TISC phases in a number of materials -- surprisingly, almost all of them are also known to display characteristics topological properties. Here we provide a review of such experiments.

\subsection{How to detect TISC with confidence?}

Since a TISC phase appears only under a point contact, the volume fraction of the superconducting phase, thus, is extremely small. In such a situation, the conventional transport and magnetization measurements technique that are used for detection of superconductivity fail.

The total resistance of a typical point contact setup can be understood by a simple model where the resistance of various components such as the resistance of the tip(R$_{Tip}$), sample(R$_{Sample}$) and the point contact(R$_{PC}$) are connected in series to form the total resistance(R$_{Total}$)(See FIG 5).   R$_{Total}$ = R$_{Tip}$ +R$_{PC}$+R$_{Sample}$.

\begin{figure}[htb]
	\centering
	\includegraphics[scale=0.4]{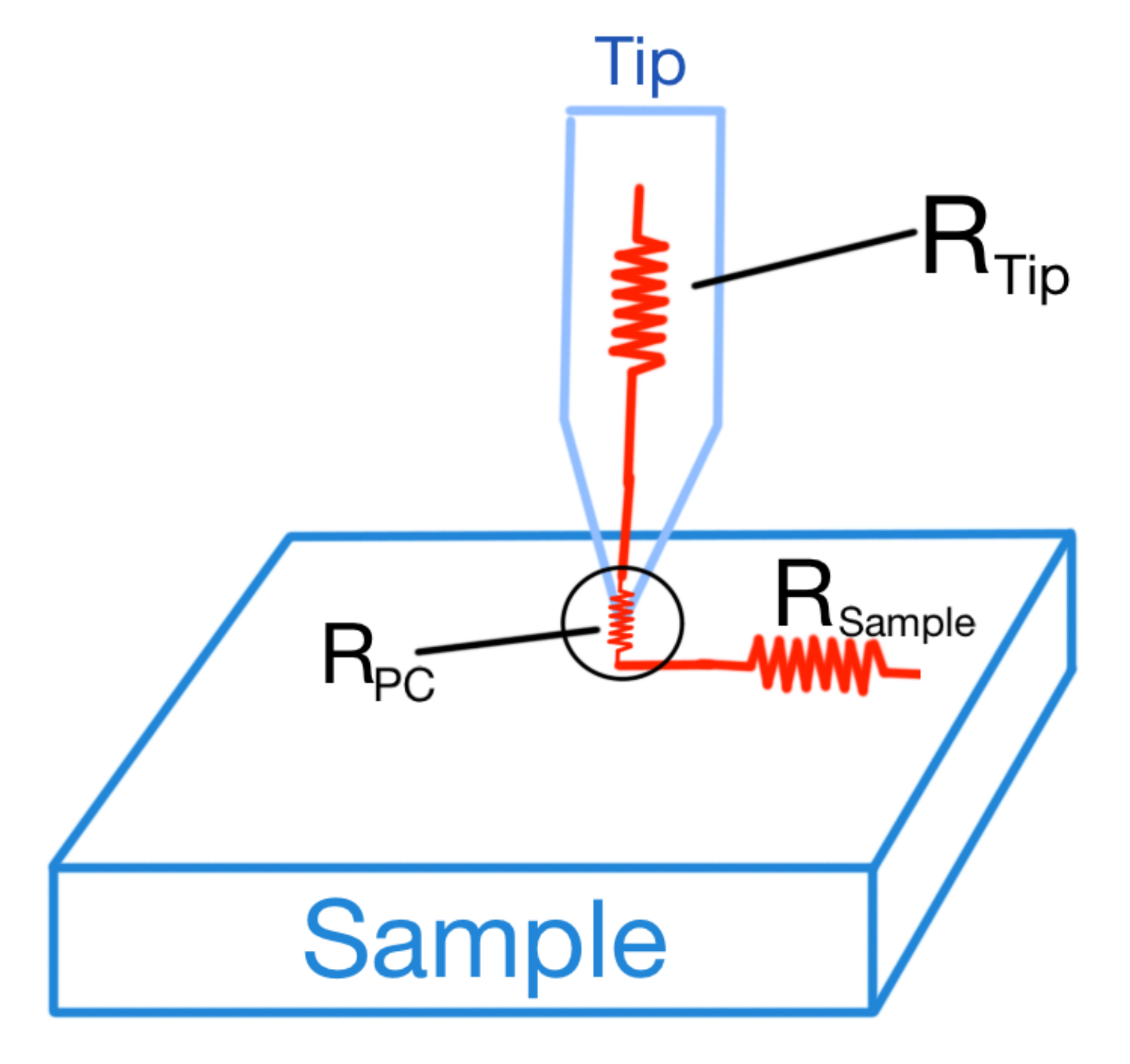}
	\caption{A schematic showing the resistance sources in a point contact geometry}
	\label{pb1}
\end{figure}

Here, one may note that even for a superconducting point contact, even when the resistance of either of the electrodes (tip or sample) is zero, the contact resistance can be non-zero. That is why, the non-dissipative state may not always be seen directly, even when the point contact is in the thermal regime of tyransport and the dominating transport channel is through the Maxwell's resistance. However, a lack of observation of the zero-resistance state does not rule out the possibility of a superconducting point contact either. As an example, we show data on point contacts with superconducting Pb (Figure 6). It can be seen that even for Pb-point contacts, the point contact resistance does not go to zero below the transition. 

A point contact interface exhibiting TISC phase is very small. The contact diameter is usually $\sim 100 nm$. Consequently, in conventional magnetization measurements, it becomes impossible to detect a tiny Meissner state that might emerge under the point contact. Such measurement would, in principle, be possible using a nano-SQUID mounted on a tip used to induce TISC.

Therefore, it is important to go beyond the conventional tools to detect TISC. Here we discuss certain experiments exploiting the point-contact geometry and the characteristic experimental signatures that could provide unambiguous evidence of superconductivity, if present, in TISC suspects. To note, while any one or two of these features may not definitively confirm TISC, all of these collectively provide conclusive signature of superconductivity.

 \begin{figure}[htb]
	\centering
 \includegraphics[scale=0.6]{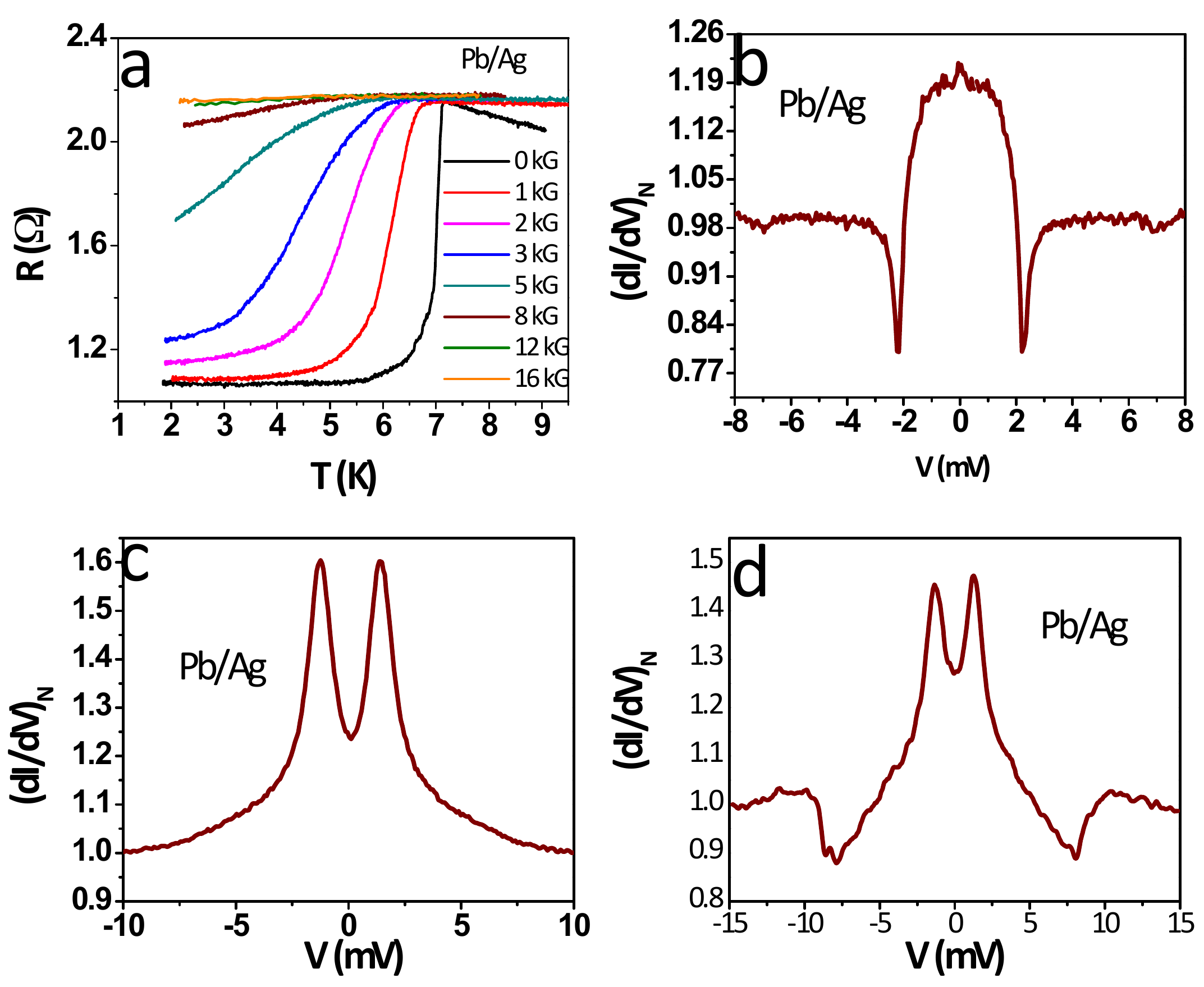}
	\caption{PCS study of superconductivity in Pb using a Ag tip. \textbf{(a)} R-T curves with systematic evolution with applied magnetic field. The $T_c$ is 7.2\,K and one may observe that the resistance in the superconducting state is finite. \textbf{(b)} Thermal regime spectra with distinct critical current dips. \textbf{(c)} Ballistic regime spectra with distinct Andreev double-peaks. \textbf{(d)} Intermediate regime spectra with critical current dips as well as Andreev peaks.}
	\label{pb1}
\end{figure}

  \textbf{a). Resistive transition in thermal regime:} In the thermal regime of point contact, the Maxwell resistance dominates and the overall resistivity of the contact is summation of resistivity due to normal component and resistivity due to superconducting component of the contact. Now if temperature dependence of such a point contact is performed then it is seen that above the critical temperature the superconducting component of resistivity becomes zero i.e the entire contact behaves as normal metal. This transition is observed as a sharp rise in the overall resistivity in the R Vs T measurement. Apart from this the sharp resistive transition evolves systematically with external applied magnetic field. The evolution is exactly similar to what is expected for a superconductor the only exception being the resistivity never going to zero below $T_c$. A typical resistance vs temperature curve of Pb/Ag point contact is shown in Fig. 6a.\\
       \textbf{b). Critical current dips in thermal regime:} If a current higher than the critical current of superconductor is passed through then the superconductors behaves as a normal metal. The same is true for superconducting point contacts where the transition is seen as appearance of sharp dips in differential conductance vs applied bias spectrum. Here, as the current through the contact is increased to $I_c$ there is sudden increase in resistance of superconductor to attain the normal state resistance value. This  leads to a huge zero bias enhancement in the differential conductance spectrum of the contact. The presence of multiple point contacts under a micro-contact leads to detection of multiple critical current dips in the dI/dV vs V spectrum. A differential conductance spectrum obtained for Pb/Ag point contact is shown in Fig. 6b. 


\textbf{c). Andreev peak in ballistic regime spectra:} Andreev reflection spectrum is obtained for superconducting point contacts in ballistic regime of transport. Such a spectrum consists of two peaks corresponding to the Andreev reflection and symmetric about the zero voltage bias at.  A representative experimental curve is shown in fig. 6c.
       \begin{figure}[htb]
	\centering
	\includegraphics[scale=0.6]{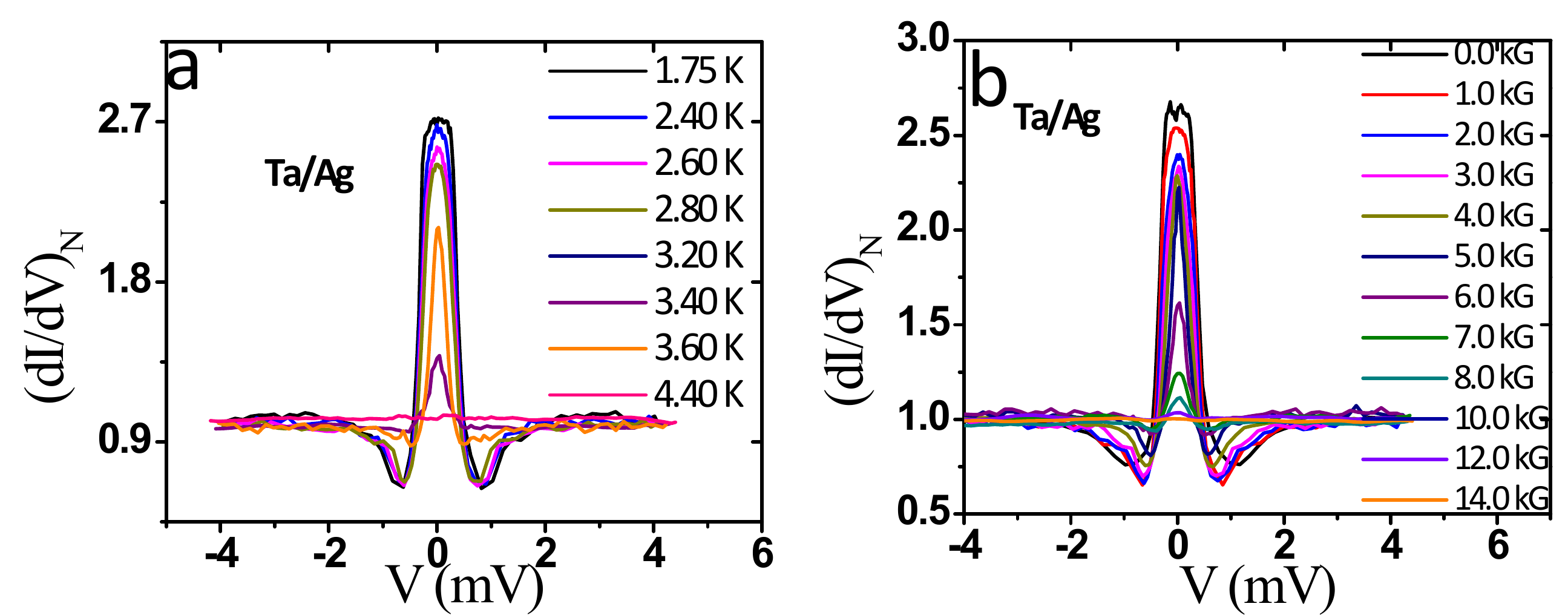}
\caption{Systematic evolution of typical thermal regime spectra on Ta/Ag point contact with variation in (a) temperature and (b) magnetic field, respectively.}
	\label{therm}
\end{figure}

\begin{figure}[htb]
	\centering
\includegraphics[scale=0.6]{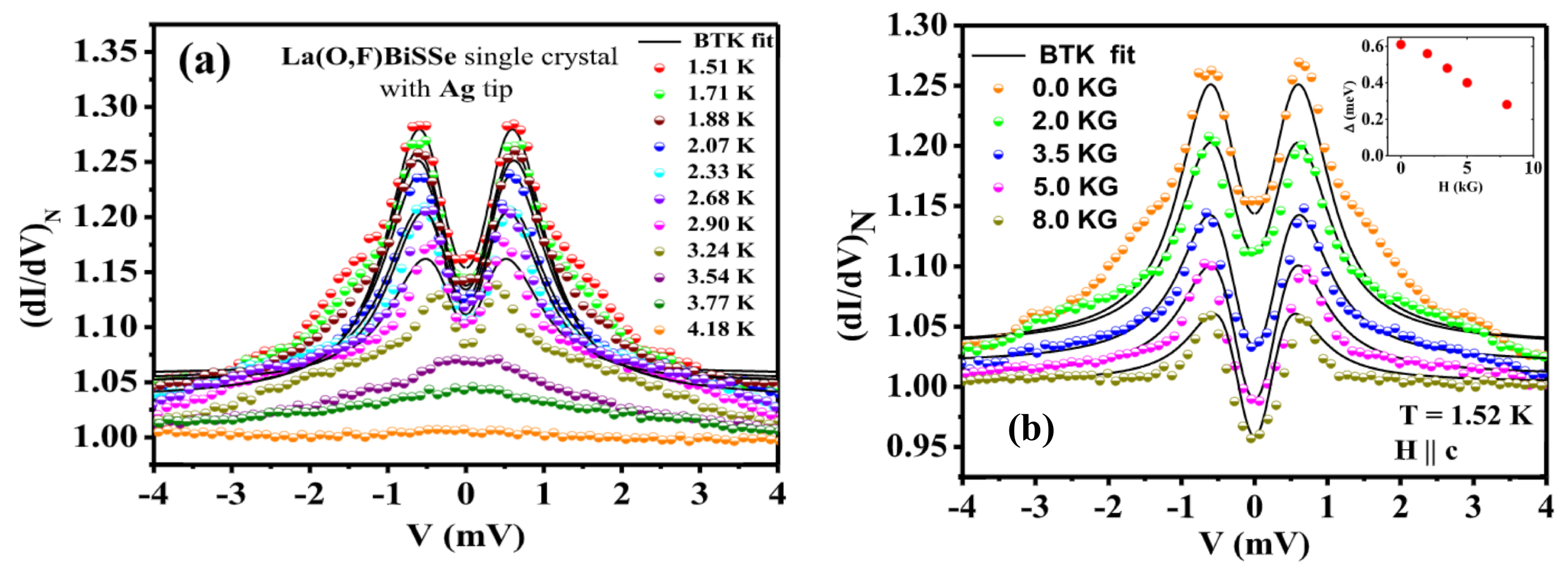}
	\caption{Systematic evolution of ballistic regime spectra obtained on La(O,F)BiSSe/Ag point contact with variation in (a) temperature and (b) magnetic field, respectively.\cite{BTK}.}
	\label{taas1}
\end{figure}
\textbf{d). Andreev peaks and critical current dips in intermediate regime:} When a  Point contact spectra show simultaneous appearance of Andreev peaks and critical current dips then it can be said that one has achieved a contact where the contact is intermediate to the ballistic and thermal contacts. Such contacts are usually realised by withdrawing the tip slowly, so that a thermal point contact transitions to an intermediate one. Energy resolved Andreev reflection spectroscopy can be performed as long as there are no inelastic scattering with in the contact region. Hence the intermediate spectra can be analysed to determine the superconducting gap energy value. An intermediate point contact spectra obtained on Pb/Ag point contact is shown in Fig. 6d.

\begin{figure}[htb]
	\centering
 \includegraphics[scale=0.6]{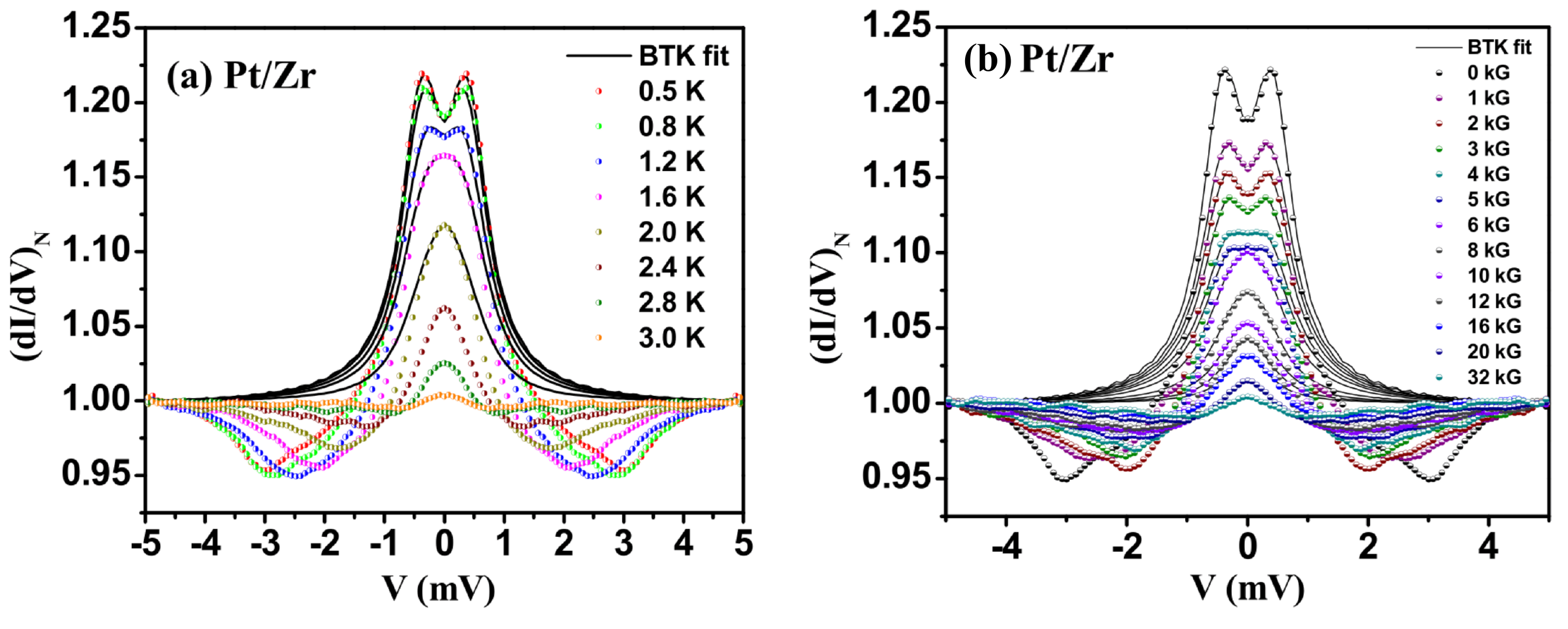}
	\caption{Systematic evolution of typical intermediate regime spectra on Zr/Pt point contact with variation in (a) temperature and (b) magnetic field, respectively.}
	\label{int}
\end{figure}

\textbf{e). Systematic evolution with H \& T:} For superconductors critical magnetic field and critical temperature are two of the most significant quantities. With tuning of external parameters such as application of magnetic fields, varying the temperature, it is expected that the spectroscopic signatures will evolve systematically and vanish at critical values of applied field and temperature. The systematic evolution helps in understanding the nature of superconductivity and provides information about the origin of superconductivity in the material. Temperature and magnetic field dependent spectra for thermal, ballistic and intermediate contacts are shown in Fig.  7,8 and 9 respectively.

\textbf{f). Observation of Shapiro steps in Josephson junction contacts and oscillations in zero-bias current with weak magnetic field:} I-V characteristics obtained for superconductor-superconductor point contacts show distinct steps when irradiated with microwave radiation\cite{Shapiro, Grimes, Zimmerman, Taguchi}. These steps are known as Shapiro steps and are exactly same as what is observed for Josephson junctions\cite{Josephson, Tinkham} This confirms that a superconducting phase is achieved under the tip where the point contact itself behaves like a Josephson junction.  Moreover, in presence of a weak magnetic field the zero bias current must exhibit oscillations. For such an experiment the applied magnetic field  is around 2a few hundreds of  milli-gauss, which is comparable to the earth's magnetic field(200-600 milli-gauss), hence special measures in the probe design are needed to to block the effect of earth's magnetic field on the experiment. 
In case of TISC, if two superconducting contacts are formed very close to each other, then effectively we obtain a superconductor/vacuum/superconductor like arrangement. An I-V measurement in presence of microwave radiation can in principle induce Shapiro steps confirming that indeed the contacts are superconducting.\\
 
Though the above tests indicate that the point contact is indeed superconducting, it is important to look at the behaviour of materials forming contact at high pressures.  Since PCAR relies on formation of contact of a sharp tip with a sample, the tip can apply few GPa of mechanical pressure on the sample. It is known that certain materials  exhibit superconducting behavior on application of pressure. For such cases due to application of local mechanical pressure by tip, superconductivity is induced in the sample. This is however not TISC but is pressure induced superconductivity. Hence, it is important to determine whether the particular sample material exhibits pressure induced superconductivity in bulk form and whether the tip can exert pressure corresponding to this value.

\subsection{Examples of TISC}
Now that we have discussed the prerequisites and salient features obtained for a system exhibiting tip induced superconductivity, it is time to discuss a few examples where this effect has been observed. In this section we start with the experimental discovery of TISC, we discuss the results in detail and then move on to other systems where TISC phase could be realized. 
\subsubsection{\textbf{TISC in Dirac semimetals (Cd$_3$As$_2$)}}

3D Dirac semimetals, in general, are known to be stable compounds due to symmetry protected surface states. Another key feature of these materials is that the Dirac points exist close to the topological phase boundaries, as a result by breaking certain symmetries they can be driven into further exotic states such as topological insulators, Weyl semimetals and topological superconductors. However, until the year 2014, this idea was not realized practically. In 2014, Dr. Goutam Sheet's group began working on a II-V semiconductor $Cd_3As_2$. The existence of a stable 3D topological semimetallic phase in $Cd_3As_2$ was previously confirmed by Angle resolved photoemission spectroscopy (ARPES) \cite{Liu}and scanning tunnelling spectroscopy(STS) \cite{Jeon} experiments. It was further observed that the energy dispersion is linear along all three directions in momentum space. The group made a very surprising discovery, the stable 3D Dirac semimetal $Cd_3As_2$ could be easily driven to a superconducting phase by simple touch with a silver(Ag) needle. The superconducting phase was found to survive upto 7.5 K with an estimated superconducting gap amplitude of 6.5 meV\cite{Aggarwal2, Aggarwal4}. Further, a thorough point contact spectroscopic study was conducted where the behaviour of the point contact was tested under different temperatures and applied magnetic field. The studies revealed that the gap energy is weakly dependent on temperature and survives upto 13 K\cite{Buchanan}. This indicates presence of a pseudo-gap and that the induced superconductivity is unconventional in nature. Although, polycrystalline samples of $Cd_3As_2$ were used in this experiment, it is important to note that the measured value of superconducting energy gap is not necessarily the result of averaging. The gap value is often  measured from a single grain due to small size of point contact. The measurement have been repeated over 100 times and every time the measured gap value turns out to be the same at a particular temperature indicating the existence of a nearly isotropic gap structure. Confirmation of this discovery was provided by Wang et. al \cite{Wang, Wang2}where they repeated the same experiment on $Cd_3As_2$ single crystals. \\

\begin{figure}[htb]
	\centering
\includegraphics[scale=0.6]{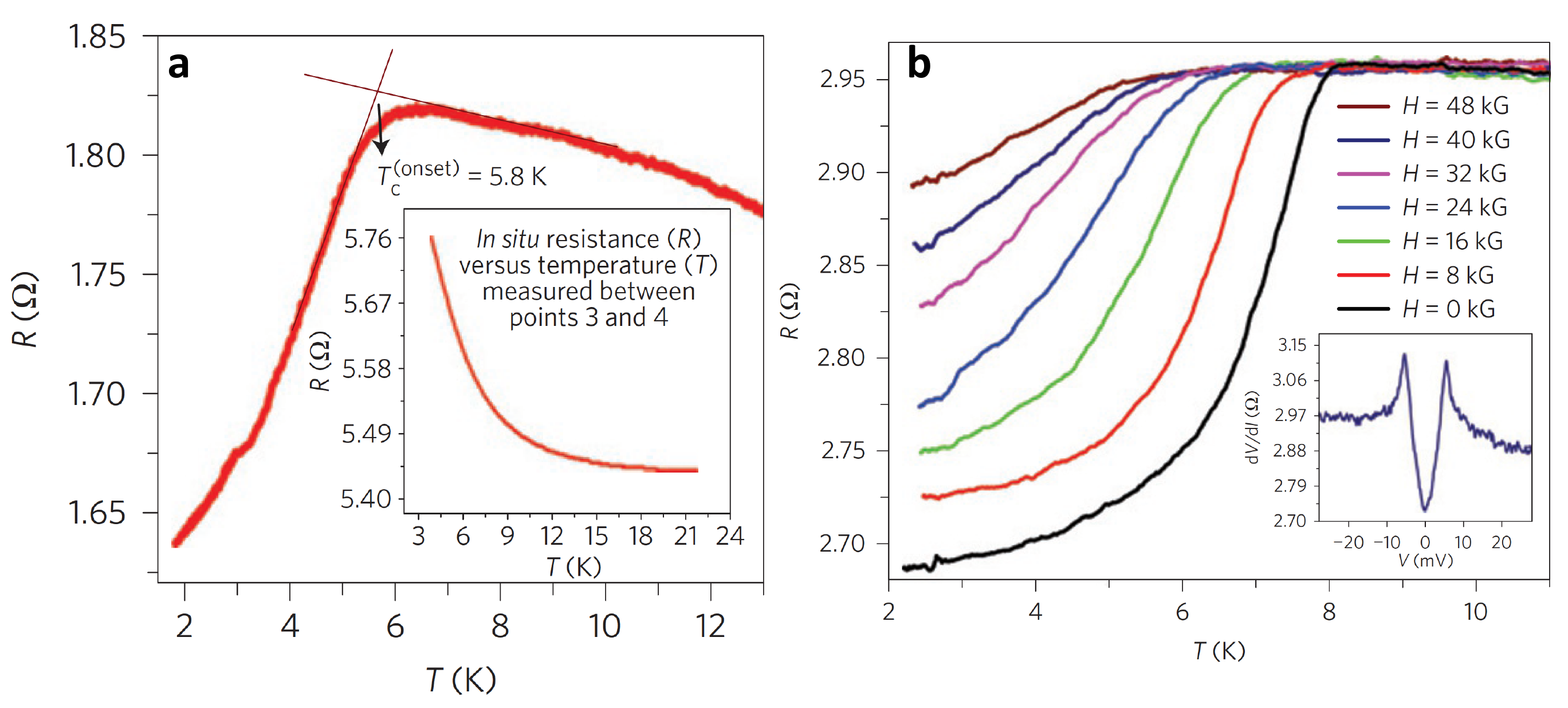}
	\caption{(a) A point-contact R-T on $Cd_3As_2$ representing superconducting transition at 5.8 K (b) H-dependence of point contact R-T showing systematic evolution of superconducting transition with applied magnetic field.}
	\label{int}
\end{figure}
\textbf{a. Point-contact spectroscopy: Different regiemes of transport on Cd$_3$As$_2$}\\
  Contacts in the thermal regime of transport for $Cd_3As_2$/Ag exhibited a sharp resistive transition around 5.8 K(Fig 10(a)). A similar transition is observed for point contacts on lead (Pb)as shown in previous sections. A magnetic field dependent study of the contact also showed a systematic decrease in the transition temperature with increasing field value (Fig. 10b) and is exactly same as expected for a superconducting point contact. Since neither the tip or the sample is superconducting alone, the emergence of a tip induced superconducting phase is indicated by this observation. A confirmation to this indication was provided by subsequent spectroscopic results.

\begin{figure}[h!]
	\centering
\includegraphics[scale=0.6]{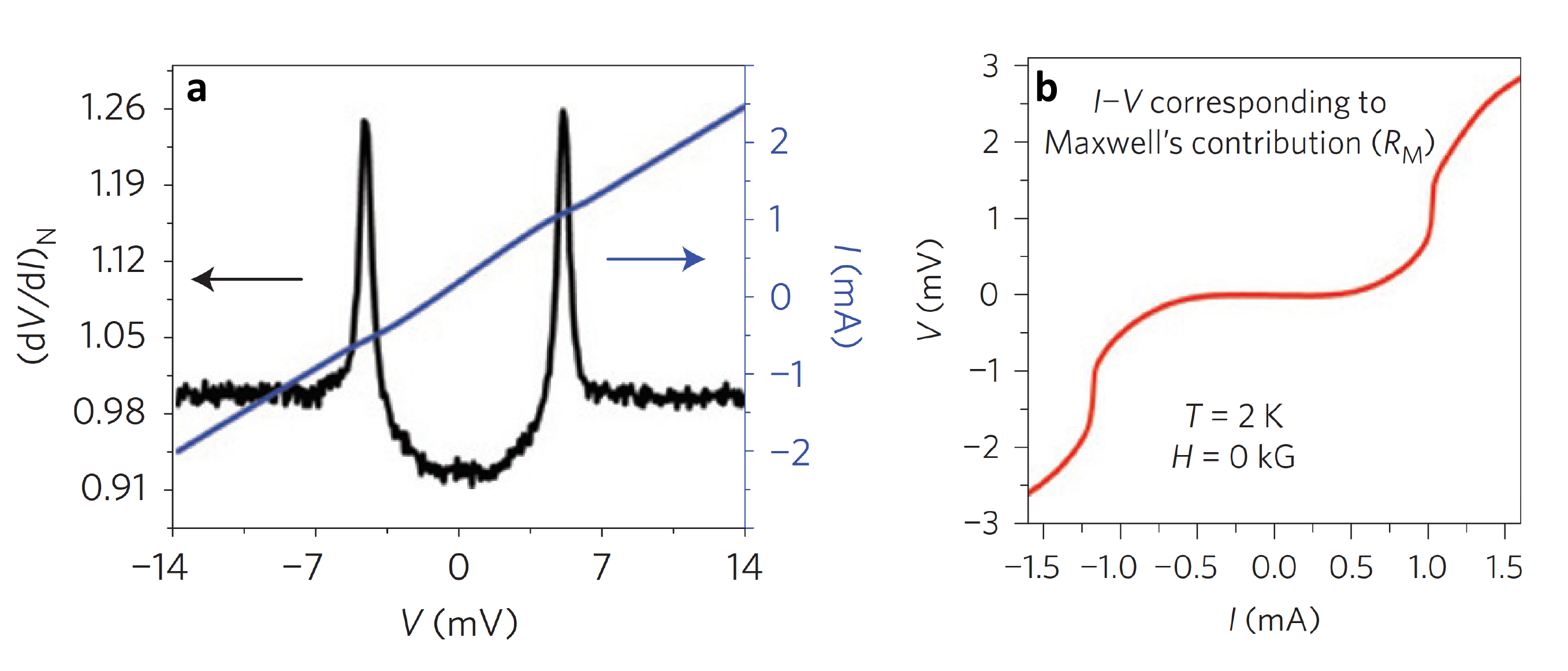}
	\caption{(a) A (dV/dI)$_N$ versus V$_(dc)$ spectrum close to the thermal regime: The peaks in (dV/dI)$_N$ originate due to the critical current of the superconducting junction. The blue line represents the I-V characteristics corresponding to the total resistance (R$_M$+R$_S$) of the point contact. (b) The I-V corresponding to R$_M$ after subtracting R$_S$ showing the dissipationless current flowing through the point-contact.}
	\label{int}
\end{figure}

\begin{figure}[h!]
	\centering
\includegraphics[scale=0.6]{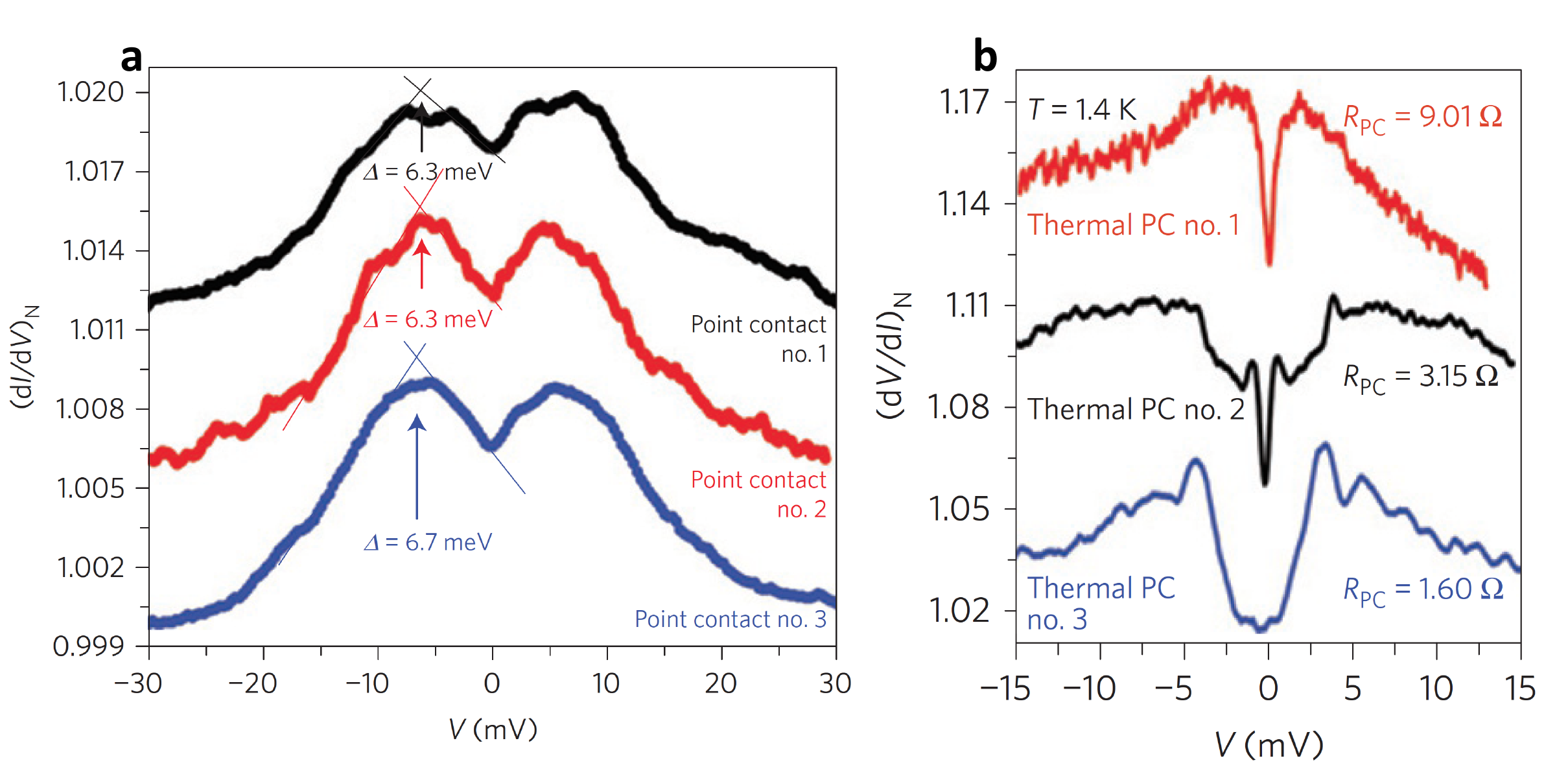}
	\caption{(a) (dI/dV)$_N$ versus V$_(dc)$ spectra in the ballistic regime of transport obtained for different  Cd$_3$As$_2$/Ag point-contacts.Here two peaks symmetric about V=0 are the AR peaks (b) (dI/dV)$_N$ vs. V spectra in thermal regime of transport at different Cd$_3$As$_2$/Ag point-contacts. These spectra have critical current driven dips .}
	\label{int}
\end{figure}

Fig 11(a) shows a normalised differential resistance spectrum $(dV/dI)_N)$ as a function of applied dc bias along with the I-V characteristics obtained for a thermal point contact between Cd$_3$As$_2$ and elemental superconductor lead (Nb). The spectrum present two sharp peaks at $\pm 5$ mV, which are typical for such point contacts \cite{Naidyuk, Sheet}.  The I-V characteristics were also calculated for contact comprising only of Sharvin resistance following the BTK theory. This I-V characteristics was then subtracted from the characteristics obtained above to obtain I-V plot corresponding to purely Maxwell contribution.  A clear dissipation less current state is seen where increase in current above the critical current value for the point contact leads to transition from superconducting state to the normal state as shown in Fig. 11(b).

The authors performed point contact experiments in different regimes of transport by manipulating the contact diameter via physically engaging/withdrawing the tip from/onto the sample surface. Fig12 (a) shows differential conductance spectrum obtained for three different contacts corresponding to  the ballistic regime of transport where, the Sharvin's resistance dominates.  The hallmark signature of Andreev reflection appear as two peaks which are  symmetric about V=0 in the spectra. \textbf{Spectra((dV/dI)$_N$ vs V$_{dc}$) corresponding to the thermal regime showing the distinct critical current dominated peaks is shown in Fig12(b)}.

Since neither the sample Cd$_3$As$_2$ nor  the tip (Silver(Ag)) are superconductors, and the spectra obtained for Cd$_3$As$_2$/Ag point contact resembles to that of a superconductor, hence it is the reasonable to compare them with spectra obtained for superconducting point contact. We find that the spectra has striking resemblance to those obtained for Pb/Ag point contact. From these results one may conclude that metallic point contacts on Cd$_3$As$_2$ are superconducting.

Since there is an indication that the point contact is superconducting, it is imperative to look at how the resistance of the contact changes with temperature. A clear transition from high resistance state to a lower resistance was observed(shown in fig. 10(a)). The transition is strikingly similar to a superconducting transition as observed for superconducting point contacts. The transition begins at 5.8 K. Now that the variation of resistance with temperature is observed, it is important to look at the dependence on magnetic field. For a superconductor it is expected that as the applied magnetic field approaches the critical field value the superconducting transition temperature decreases and just above the critical field the material remains in its normal state. As expected, the resistivity plot evolves with the applied field and the at applied filed of 4.5 T the system only exhibits normal resistance state (shown in FIg. 10(b)). 
\begin{figure}[htb]
	\centering
\includegraphics[scale=0.6]{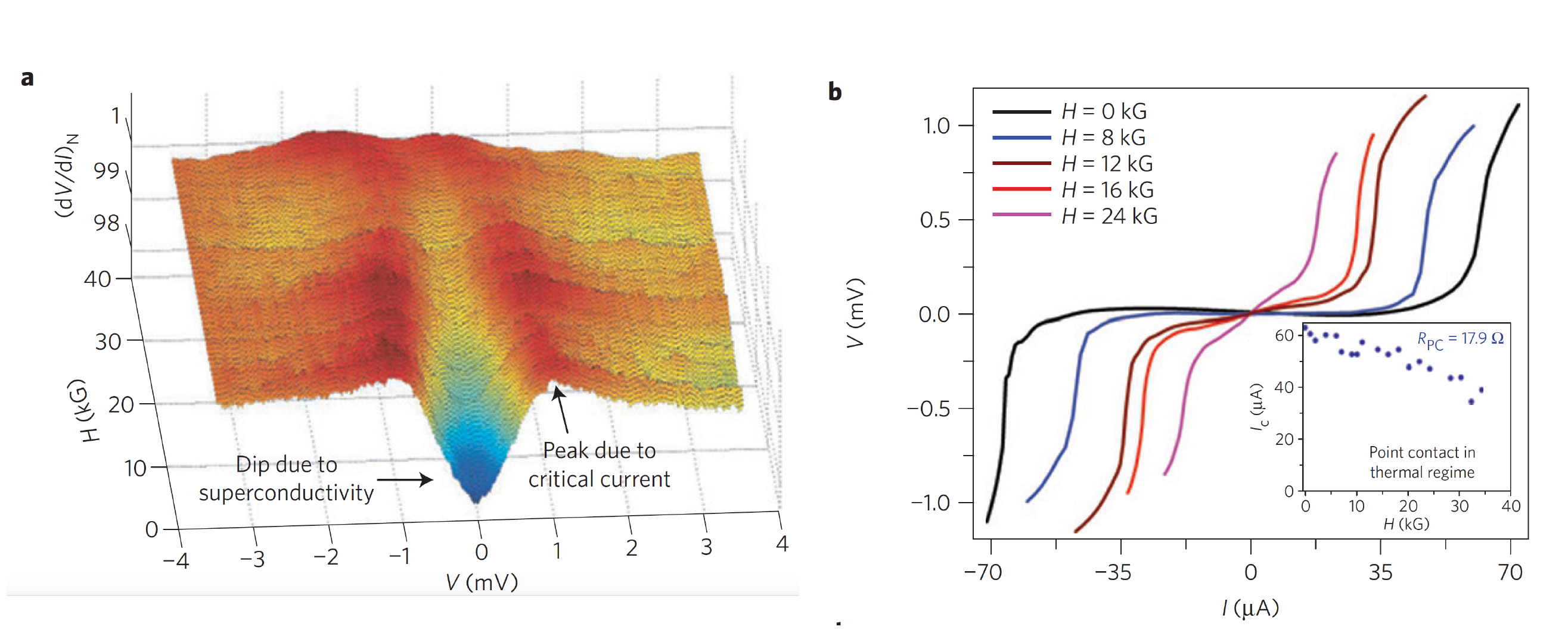}
	\caption{(a) Field dependence of  dI/dV spectra with critical current dominated peaks. (b) Corresponding  field dependence of I-V spectrum of the same contact in thermal  regime of transport}
	\label{int}
\end{figure}
Now we focus on the  effect of field on differential resistance spectrum and provide an estimation of point contact size.\\
\textbf{a. Magnetic field dependence}: Fig.13(a) shows the evolution of a critical current dominated spectra obtained in the thermal limit with applied magnetic field \cite{Sheet}. The central dip is a result of superconducting transition below the critical current. The spectra evolved smoothly with field and the peaks and dips disappear at 4.5 T.  The magnetic field dependence of the critical current dominated part of the I-V characteristics corresponding to  R$_M$ is shown in fig. 13(b). The position of the peaks in the resistance spectrum gives the approximate value of the critical current for a given point-contact (inset of fig.13(b)). As expected for superconductors, the critical current decreases with increasing magnetic field.

\textbf{b. Estimation of the point-contact size}. The  contact size can be estimated from the normal state resistance i.e at high voltage bias by the wexler's formula [$R_{PC}=\frac{2h/e^2}{(aK_f)^2} + \Gamma(l/a)\frac{\rho(T)}{2a}$] where, $\Gamma(l/a)$ is a numerical factor close to unity, a is the contact diameter and 2h/e$^2$ is quantum of resistance i.e. 50 kOhm , k$_f$ is the magnitude of the Fermi wave vector which is 0.04  $\AA$ for Cd$_3$As$_2$ and $\rho (T)$ is resistivity at any given temperature T.  Estimation of contact diameter requires the contact to be approximated as circular in nature. The value of $\rho$ at 1.5 K for Cd$_3$As$_2$ as measured by conventional four-probe method is 28  ohm. By using this value the point-contacts are estimated to be approximately  40-400 nm for point-contact resistance R$_{PC} $ 2-200 Ohm.\\

\textbf{B. Unconventional Nature of Superconductivity}

Now that it has been established that the Cd$_3$As$_2$/Ag point-contact indeed behaves as superconductor we shift our focus towards the nature of superconductivity. 
\begin{figure}[h!]
	\centering
\includegraphics[scale=0.5]{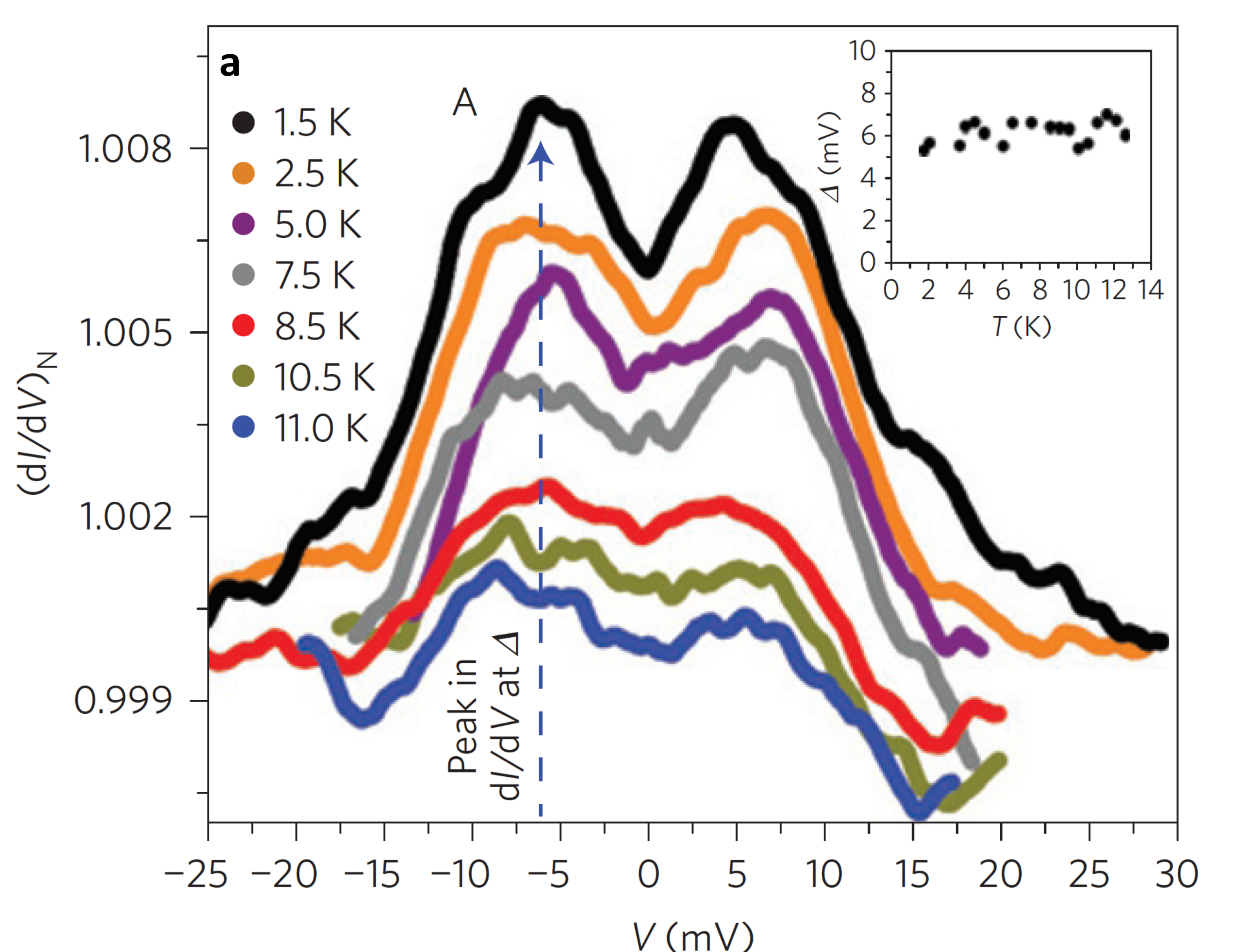}
	\caption{ Variation of differential conductance spectrum obtained in ballistic regime of transport with applied magnetic field.}
	\label{int}
\end{figure}
\textbf{a. Existence of Pseudogap} The determination of nature of superconductivity requires spectroscopic investigation of a point-contact in absence of inelastic scattering in the contact region namely the ballistic regime of transport\cite{Naidyuk}.  A  differential conductance spectrum with distinct Andreev peaks (hallmark signature of Andreev reflection)\cite{BTK} is shown in fig. 14. The Andreev peaks appear at 6.7 meV and the observed peaks were broader that what one expects from BTK theory used for analysing the AR spectra obtained for BCS superconductors\cite{Bardeen1, Bardeen2}. The broadening could be the result due to large inelastic broadening\cite{Plecenik} parameter at the interface and/or due to an unconventional pairing. The quantitative analysis of such spectra requires a theoretical model for AR which includes the non-trivial topological properties of one of the electrodes forming the point-contact. Nonetheless, an approximate estimate of the gap\cite{BTK} can be obtained via the position of  Andreev peaks. The gap is approximately 6.7 meV, an unusually large value given a low Tc $> $ 8K. Such  values for superconducting phase points to abnormally large value for $\Delta/K_BT_C \sim 10$.

Moving on to the variation of ballistic point-contact spectra with temperature, it was found that the position of Andreev peaks in the conductance spectra did not show any significant shift with increase in temperature. The peaks survive upto 13 K above which, the spectra become flat due to thermal broadening. This observation is strikingly similar to the pseudo-gap feature observed in case of the cuprates where the peaks in (dI/dV)$_N$ did not shift with temperature\cite{Buchanan,Renner}. It is evident from temperature dependence of ballistic spectra presented in Fig. 14 that  the onset temperature of superconductivity is 6K, however, the AR like features survive upto 13K. Similar observations have been made previously in ferro-pnictide\cite{Sheet2, Arham} and chacogenide superconductors\cite{Arham}. There are two possibilities which may give rise to pseudo gap in superconductors: 1) features similar to AR can originate from pre-formed phase incoherent Cooper pairs\cite{Choi} in normal state of epitaxial thin films of ferro-pnictide superconductors\cite{Sheet2} and (2) a novel phase of matter which is unrelated to superconductivity could appear in the normal state of iron chalcogenide superconductors\cite{Arham}. However, in the latter case a systematic temperature dependence of spectra features was observed unlike in the case of Cd$_3$As$_2$.

\begin{figure}[htb]
	\centering
	\includegraphics[scale=0.45]{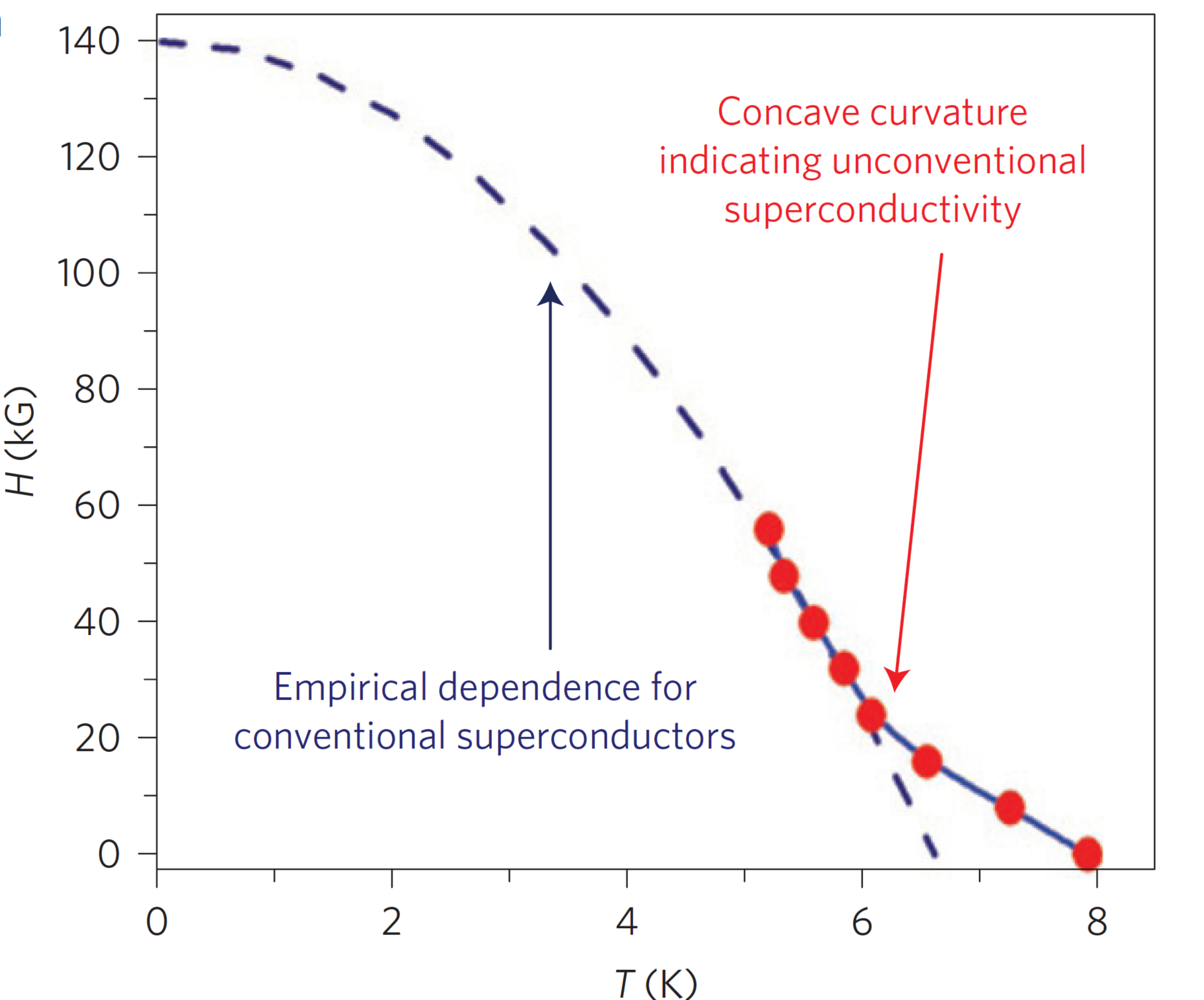}
	\caption{H-T phase diagram constructed from field dependence of Resistivity measurements along with an empirical H-T prediction.}
	\label{int}
\end{figure}
\textbf{b. Concave curvature of H-T phase diagram}:  As we have discussed before, the variation of R-T curves with applied magnetic field(Fig.10(b)) revealed that the T$_c$ decreases with increasing magnetic field. A H-T phase diagram (fig. 15) was constructed and it was compared with predicted curve for a conventional superconductor. An approximate extrapolation of the empirical H-T curve indicated a high critical field of 14 T. Deviation of experimental curve from empirical prediction at higher temperature with a concave curve indicated towards unconventional superconductivity\cite{Tinkham}. However, concave curvature of H-T phase diagram can arise from any intrinsic disorder induced by the point contact. 

\textbf{c. Zero Bias Conductance Peak(ZBCP:)} Upon further investigation of the nature of superconductivity and associated gap amplitude via magnetic field dependence of a ballistic spectrum it was found that the gap features show systematic evolution with magnetic field. The extrapolation of this dependence indicated that the gap vanishes around 45 kG implying that the pseudo-gap in this case is a precursor phase to the superconducting state.

\begin{figure}[htb]
	\centering
	\includegraphics[scale=0.6]{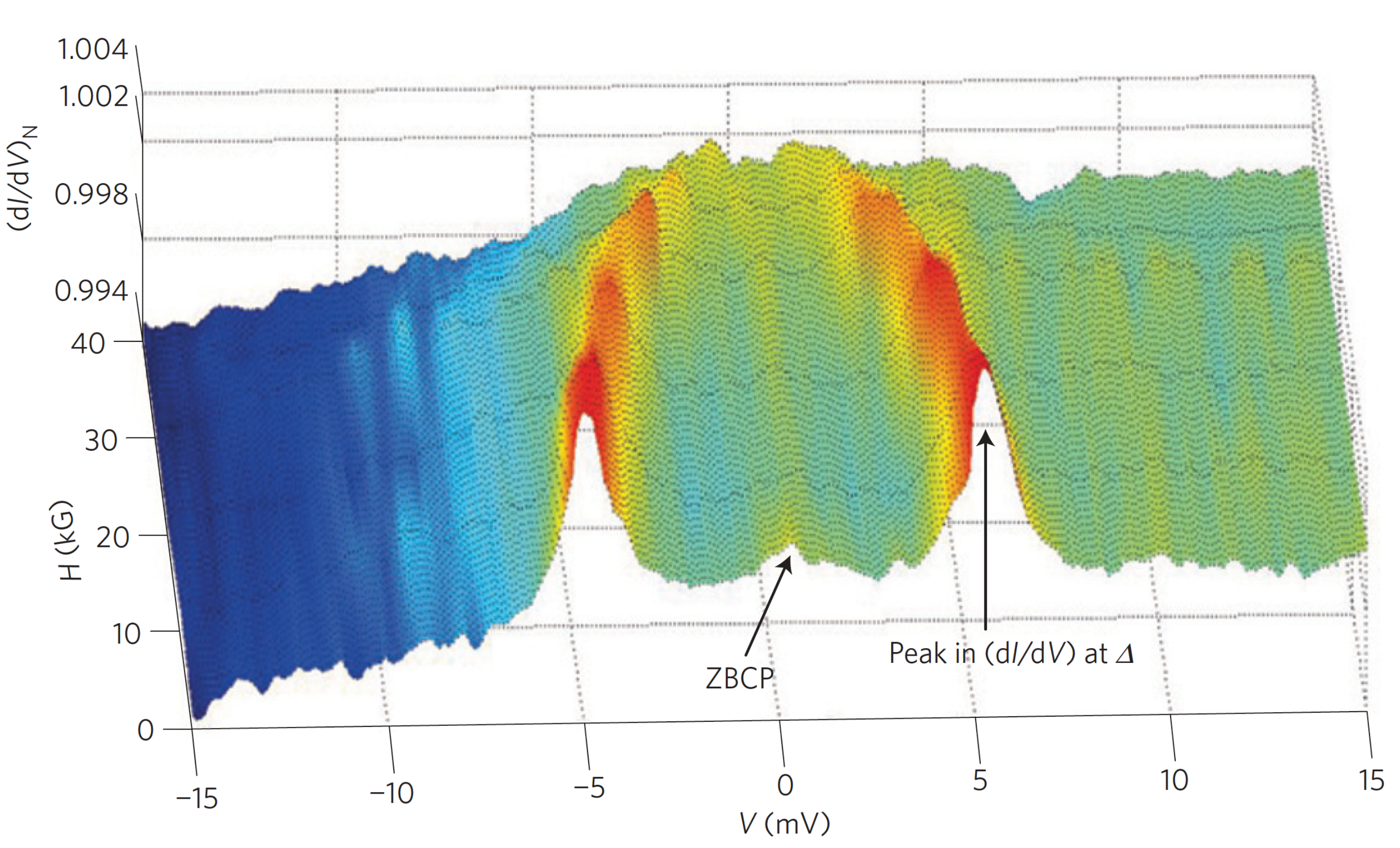}
	\caption{Magnetic field dependence of a spectrum obtained in the ballistic regime showing the evolution of the gap feature as well as a prominent ZBCP with magnetic field.}
	\label{int}
\end{figure}

For the Cd$_3$As$_2$/Ag point-contacts, all the differential conductance spectra obtained in the ballistic regime of transport showed a peak structure(for large overall signal) at V=0. For certain point contacts, the authors obtained pronounced zero bias conductance peak(ZBCP). Such pronounced ZBCP(Fig. 16) is commonly seen for ab-plane tunneling in d-wave superconductors where, the ZBCP appear due to existence of zero energy Andreev bound states(ABS)\cite{Tanaka}. For d-wave junctions the ZBCP undergo splitting upon application of external field\cite{Tanaka1}. However, in the presented case, the authors did not observe any splitting of ZBCP rather the peak slowly faded away with increasing field. This unique field dependence hints at a possible p-wave component in the order parameter symmetry of new superconducting phase giving rise to the zero energy Andreev bound states\cite{Tanaka1}. Moreover, the field dependence established that the ABS is robust and can survive high magnetic fields. Here, the superconducting phase was derived from a topologically non-trivial system hence, the robust nature of ABS against large magnetic field is also indicator of existence of time-reversal invariant Majorana edge-modes in Cd$_3$As$_2$ point contacts\cite{Sato1, Tanaka2}.  A similar ZBCP with similar dependence with magnetic field was observed in topological superconductor Cu$_x$Bi$_2$Se$_3$ and the ZBCP was attributed  to existence of Majorana Fermions.\cite{Sasaki, Takami, Yamakage}

\textbf{d. Order parameter symmetry} A visual inspection of the dV/dI spectra presented in fig.18 reveals that the spectra has contributions from both AR and critical current and belongs to the intermediate regime.  A spectrum with only Andreev peaks(Ballistic regime) can be fitted perfectly under the BTK theory to obtain information about the gap energy, barrier potential etc.  In the ballistic regime, the inelastic scattering processes hence in the case of Cd$_3$As$_2$/Ag point contacts , the unconventional component of the order parameter are prominent resulting in spectra which is significantly broader than the BTK prediction.

The authors were able to fit the low-bias portion of the intermediate spectra with the BTK thory. This indicated that the order parameter symmetry in the new superconducting phase emerging at Cd$_3$As$_2$/Ag point contact has a mixed angular momentum symmetry with a strong s-wave component. Moreover, the analysis of the observed ZBCP also provides strong indication towards the possibility of $s+p$-wave type of symmetry in this new superconducting phase\cite{Kashiwaya}.



\textit{\textbf{Splitting of ZBCP in Cd$_3$As$_2$/Nb point-contact}}

The investigation of superconducting systems in complex systems such as topologically non-trivial systems like Cd$_3$As$_2$ \cite{Liu, Young, Liu2} has provided the means to explore the topological superconducting phase hosting the elusive Majorana fermions\cite{Wilczek1, Wilczek2}. A significant progress has been made, from a theoretical perspective, in understanding the possible signature of Majorana modes emerging at the interfaces between a known s-wave superconductor and a high spin-orbit coupled semiconductor\cite{Ando1, Hui, Cole, Setiawan, Takei, Liu5, Stanescu, Lutchyn, Kim, Hui1, Ganeshan, Stanescu1}. Point contact spectroscopic studies on  such system  with a conventional s-wave superconductor  may enable us to explore such possibilities. Here, the authors performed point contact spectroscopy on Cd$_3$As$_2$ using a conventional s-wave superconductor Niobium(Nb). A  schematic showing the formation of point contact is shown in Fig17 (a).

\begin{figure}[htb]
	\centering
\includegraphics[scale=0.5]{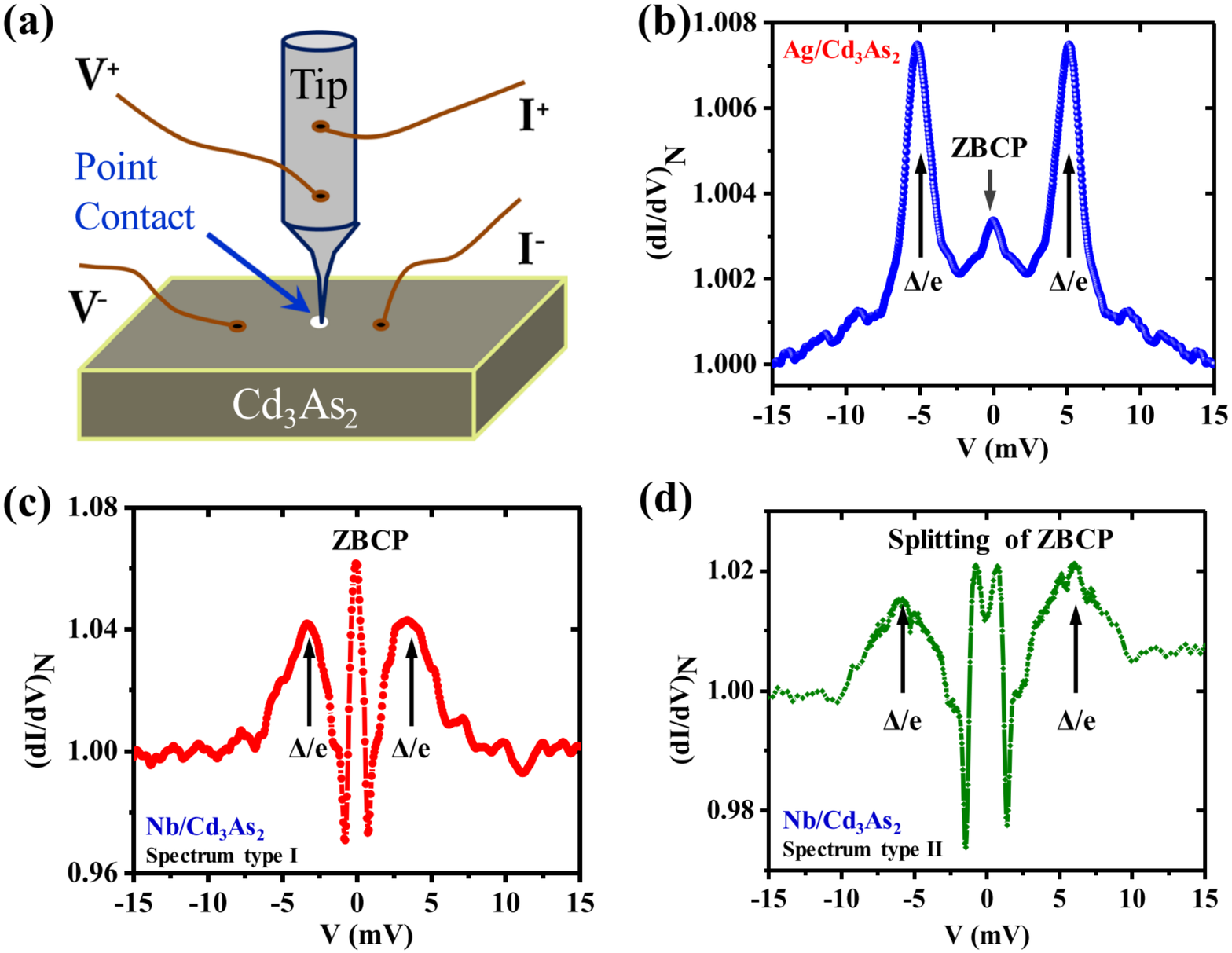}
	\caption{(a) A representative needle anvil point contact between Cd$_3$As$_2$ and Niobium (Nb). (b)A spectrum obtained for Ag/Cd$_3$As$_2$ point-contact showing zero bias conductance peak(ZBCP) along with two sharp peaks corresponding to the induced superconducting gap voltage. (c) A Spectrum obtained for Nb/Cd$_3$As$_2$ point contact showing a very prominent ZBCP without splitting- we call this type-I spectrum. (d) Spectrum obtained for a different barrier strength where the ZBCP shows splitting-we call this type-II spectrum}
	\label{int}
\end{figure}
\newpage

\textbf{a. Point-contact spectroscopy in different regimes of transport using Nb tip:} Typical point contact spectra are shown in fig. 17. It is clearly seen in the fig. 17(b) that a small zero-bias conductance peak (ZBCP) is  observed in differential conductance spectrum obtained for Cd$_3$As$_2$/Ag point contacts. However, a visual inspection of the spectrum presented in fig. 17(c) reveals that the ZBCP observed for Cd$_3$As$_2$/Nb point contacts are remarkably sharper to those obtained for Cd$_3$As$_2$/Ag point-contacts. Furthermore, it was observed that as the point-contact is physically altered to achieve a different  barrier potential (Z, as defined by the BTK theory), the ZBCP spontaneously split into two peaks symmetric about V=0 (fig.17(d)). The point contact experiments were performed using Nb tips made from wires of pure(99.9999\%) Nb with cross-sectional diameter of 0.25 mm. The wires had a superconducting critical temperature of 9.1 K with a superconducting energy gap of 1.2 meV. The spectra obtained for Nb/Cd$_3$As$_2$ were broader than Ag/Cd$_3$As$_2$ point contacts. The broadening of spectra in this case can be attributed to non-trivial mixing of s-wave gap with the unconventional gap of Cd$_3$As$_2$ and results in masking of AR features associated with Nb.

\begin{figure}[h!]
	\centering
\includegraphics[scale=0.45]{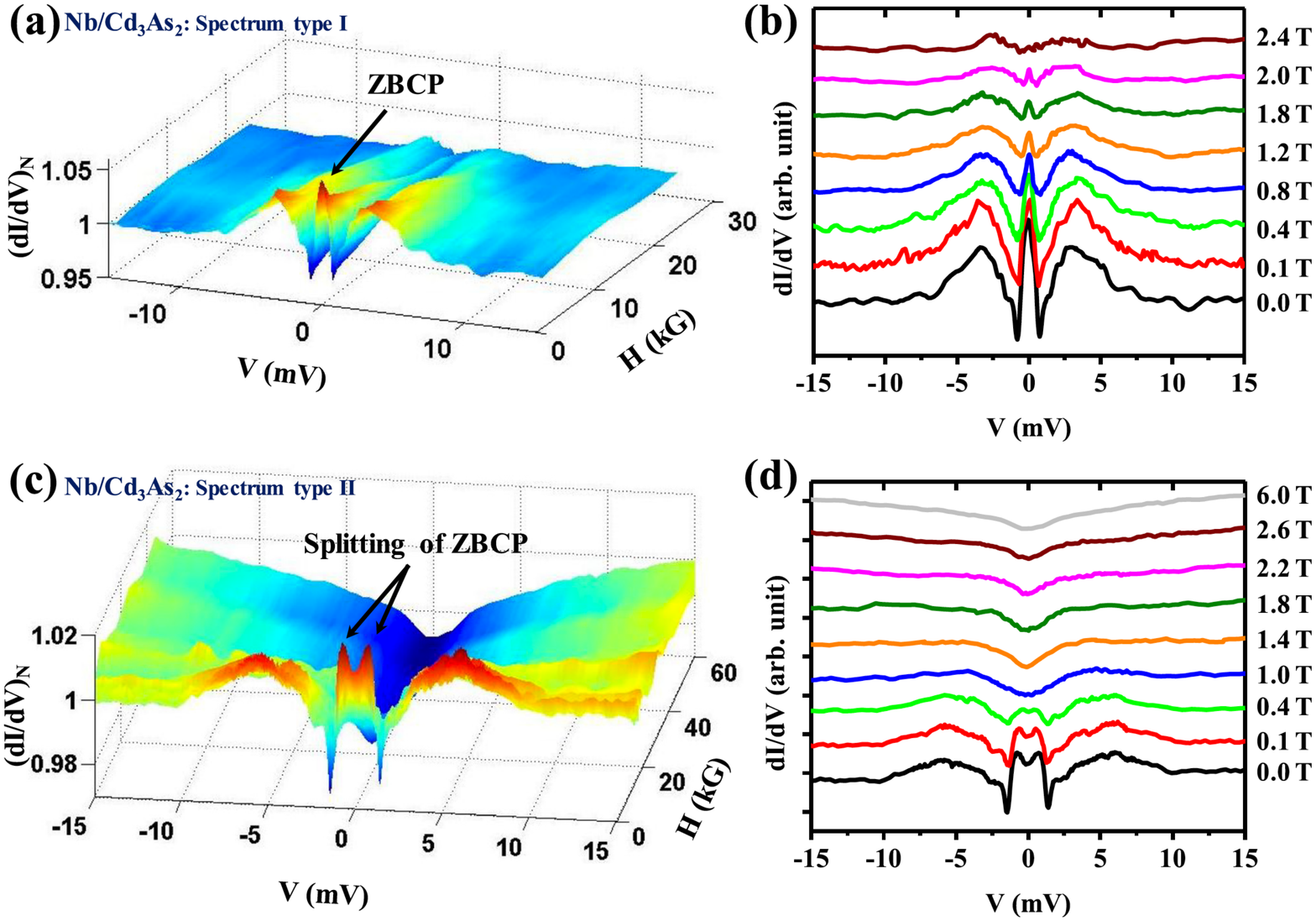}
	\caption{(a),(b) Magnetic field dependence of a type-I spectra obtained for Nb/Cd$_3$As$_2$ point-contact in 3-D and 2-D representation respectively. (c),(d) Magnetic field dependence of type-II spectra obtained for Nb/Cd$_3$As$_2$ point-contact in 3-D and 2-D representation respectively. The point-contact spectra shows disappearance of a ZBCP and splitting of the ZBCP with stronger magnetic field.}
	\label{int}
\end{figure}
    A systematic, smooth evolution of ZBCP with magnetic field for type I ( prominent ZBCP without splitting obtained for Nb/Cd$_3$As$_2$) spectra and type II ( ZBCP with splitting obtained for Nb/Cd$_3$As$_2$ for a different barrier strength) spectra is presented in fig. 17.

Type I spectra loses its prominent spectral features -the induced gap structure at  5 meV and the ZBCP at an applied magnetic field of 2.4 T (fig. 18(a),(b)). It can be clearly seen that the ZBCP and the induced superconducting gap structure disappears at the same magnetic field. This implies that the observed ZBCP is related to superconductivity. A similar field dependent behaviour of the ZBCP was demonstrated earlier with non-superconducting tips. A type-II spectra that show splitting of ZBCP also evolve systematically with increasing magnetic field(Fig18(c,d)). The double peak structure transforms into a single peak at 0.4 T which then further evolves to become a single dip at 1.0 T. This dip along with the induced gap structure disappears at a high magnetic field of 6 T. The observed difference between the magnetic field at which the the spectral features disappear for the type I and type II spectra indicated a significant difference between the two types of point-contacts. Furthermore, the observed difference in field dependence for the two types of points contacts  is consistent with expected variation. 

\begin{figure}[h!]
	\centering
\includegraphics[scale=0.5]{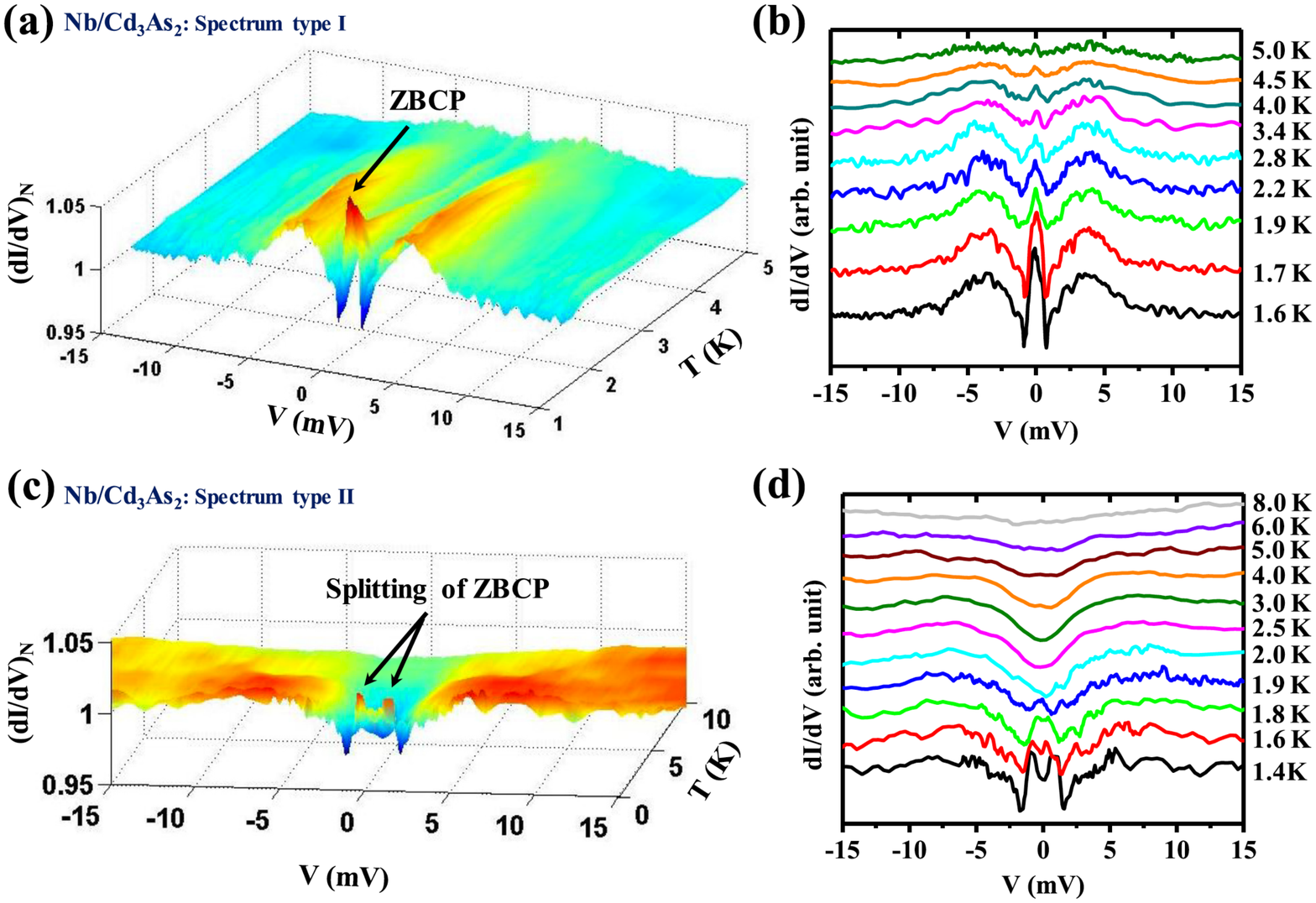}
	\caption{(a),(b) Temperature dependence of a type-I spectra obtained for Nb/Cd$_3$As$_2$ point-contact in 3-D and 2-D representation respectively. (c),(d) Temperature dependence of type-II spectra obtained for Nb/Cd$_3$As$_2$ point-contact in 3-D and 2-D representation respectively. The point-contact spectra shows disappearance of a ZBCP and splitting of the ZBCP with higher temperature.}
	\label{int}
\end{figure}

The temperature dependence of AR spectra of type I and type II also confirmed the relation between ZBCP and its splitting(see fig. 19). In this case, the spectral features again evolved systematically with increasing temperature and the ZBCP disappears along with the superconducting features at the same temperature. Moreover, it was observed that the critical temperature of the point contacts vary slightly between the type-I and type-II spectra probably due to the difference in the contact geometry and transparency of contacts. 

Though Cd$_3$As$_2$ is not a bulk superconductor, the presented observations and results strongly indicate the presence of unconventional order parameter in mesoscopic contacts of Cd$_3$As$_2$ and hence unconventional superconductivity. However, to establish the exact nature of order parameter symmetry and the underlying pairing mechanism  it is necessary to perform further theoretical modeling \& analysis as well as additional experimental investigations such as planar tunnelling, high-pressure measurements etc. 
The point-contacts on Cd$_3$As$_2$ with other metallic non superconducting tips such as Platinum(Pt) and Gold(Au) also exhibit superconductivity with properties similar to what has been previously observed with Silver(Ag). By investigating the key features  of complex band structure of Cd$_3$As$_2$ might help in understanding the aspects of emergence of local exotic superconducting phase. A first-principle calculations showed that Cd$_3$As$_2$ is a symmetry-protected topological semi-metal. The bulk of Cd$_3$As$_2$ has a pair of 3D Dirac points at the Fermi level and the surfaces have Fermi arcs\cite{Wang2}. In this context, one can possibly drive Cd$_3$As$_2$ into topologically distinct phases like a Weyl semi-metal, a topological insulator or topological superconductor by breaking certain symmetries associated with the system. There are two possible mechanisms responsible for emerging superconducting properties in the point-contacts formed on Cd$_3$As$_2$: (a) The metallic tip can present conditions favourable for local superconductivity by acting like a local dopant and modifying the local carrier concentration\cite{Liu}. (b) the formation of point contact could alter the local band structure just below the tip due to physical alteration in the local structure and lowering of symmetry rhelps in stabilizing of a local superconducting phase\cite{Kobayashi}. However,  the origin of this superconducting phase remains elusive and requires further theoretical investigations.

\subsubsection{\textbf{TISC in Weyl semimetals (TaAs,TaP etc.)}}

\textbf{ TaAs:}

The discovery of Weyl semimetals\cite{Wan, Burkov, Xu6, Lv, Xu7, Huang1, Lv1, Huang2, Weng3, Zhang2, Shekhar} facilitated the realization of Weyl fermions in condensed matter systems after more than 80 years from their theoretical discovery\cite{Xu8}. Weyl fermions were first shown by Hermann Weyl to emerge as solutions to the relativistic Dirac equation\cite{Xu6, Zhang2} in quantum field theory. Weyl fermions remained illusive in nature until the discovery of TaAs as a Weyl semimetal\cite{Xu6, Lv,Huang1, Lv1, Huang2, Weng3, Zhang2}. Weyl semimetals belong to the class of topologically non-trivial materials known to demonstrate exotic quantum phenomena having unique surface states\cite{Zhang3}. The bandstructure of Weyl semimetals contain a pair of Weyl nodes at the Fermi level where each node,in the momentum space,  can be considered as a monopole/anti-monopole of Berry curvature\cite{Lv, Murakawa}. Each Weyl node is associated with a quantized Chiral charge  and the Weyl nodes are connected to eachother through  the boundary of the crystals through Fermi arcs\cite{Xu6, Huang1, Lv1, Zhang2,Chang}. Recent investigations of Weyl semimetal TaAs have revealed existence of highly spin polarized Fermi arcs lying in a completely 2D plane on the surface of the crystal\cite{Xu8}. Weyl semimetals are believed to posses a richer set of physical phenomena stemming from its exotic topological properties presenting a system which must be explored thoroughly (a) for gaining insight into the world of quantum mechanics (b)  to find potential device applications. In this direction TISC phase realized on TaAs\cite{Aggarwal1, Gayen1} is a significant step forward.\\
In the subsequent section, the emergence of a tip-induced superconducting phase in mesoscopic contacts between silver (Ag) and a Weyl semimetal is discussed in detail pertaining to transport and magneto transport measurements  in various regimes of mesoscopic transport. The experimental data obtained for Ag/TaAs point contacts were fitted under the modified BTK formalism that includes spin polarization as a fitting parameter. The experiments and fitting pointed towards indicated the coexistence of superconductivity and high transport spin polarization  and hence indicated a flow of spin polarized super-current through TaAs point contacts. The TISC phase obtained in TaAs system did not show the possibility of an unconventional pairing mechanism, the spectra obtained in the ballistic or intermediate regime of transport did not show any zero-bias conductance peak or any features resembling the pseudo-gap. This is in clear contrast to the TISC phase obtained on the 3D Dirac semimetal Cd$_3$As$_2$. The unique co-existence of superconductivity and high transport spin polarization in TaAs point contacts makes it an interesting candidate for spintronic applications.\\

\textbf{a. Point contact spectroscopy in different transport regimes on TaAs}

Point contact spectroscopy was performed on single crystals of TaAs using elemental silver (Ag) tip. Fig. 20 (a) shows a spectra obtained in the thermal regime of transport where a broad conductance peak can be seen along with critical current dominated dips indicating the existence of a superconducting phase\cite{Naidyuk, Sheet}. A temperature dependent study of the point contact resistance shows systematic evolution with applied magnetic field, presented in fig.20(b), consistent with expected signatures of superconductivity.  The transition from normal state to the superconducting state occurred at 7.3 K in absence of magnetic field. With application of field and increase in magnitude, as expected for materials exhibiting superconductivity, the transition temperature kept decreasing with increasing magnetic field. Though the above data hinted at existence of a superconducting phase in Ag/TaAs point contacts, additional data is required to in support of the superconducting phase. In this regard, the existence of superconductivity must be established by driving the point contact into the other regimes of transport where  features related to AR can be observed. 

\begin{figure}[h!]
	\centering
	\includegraphics[scale=0.5]{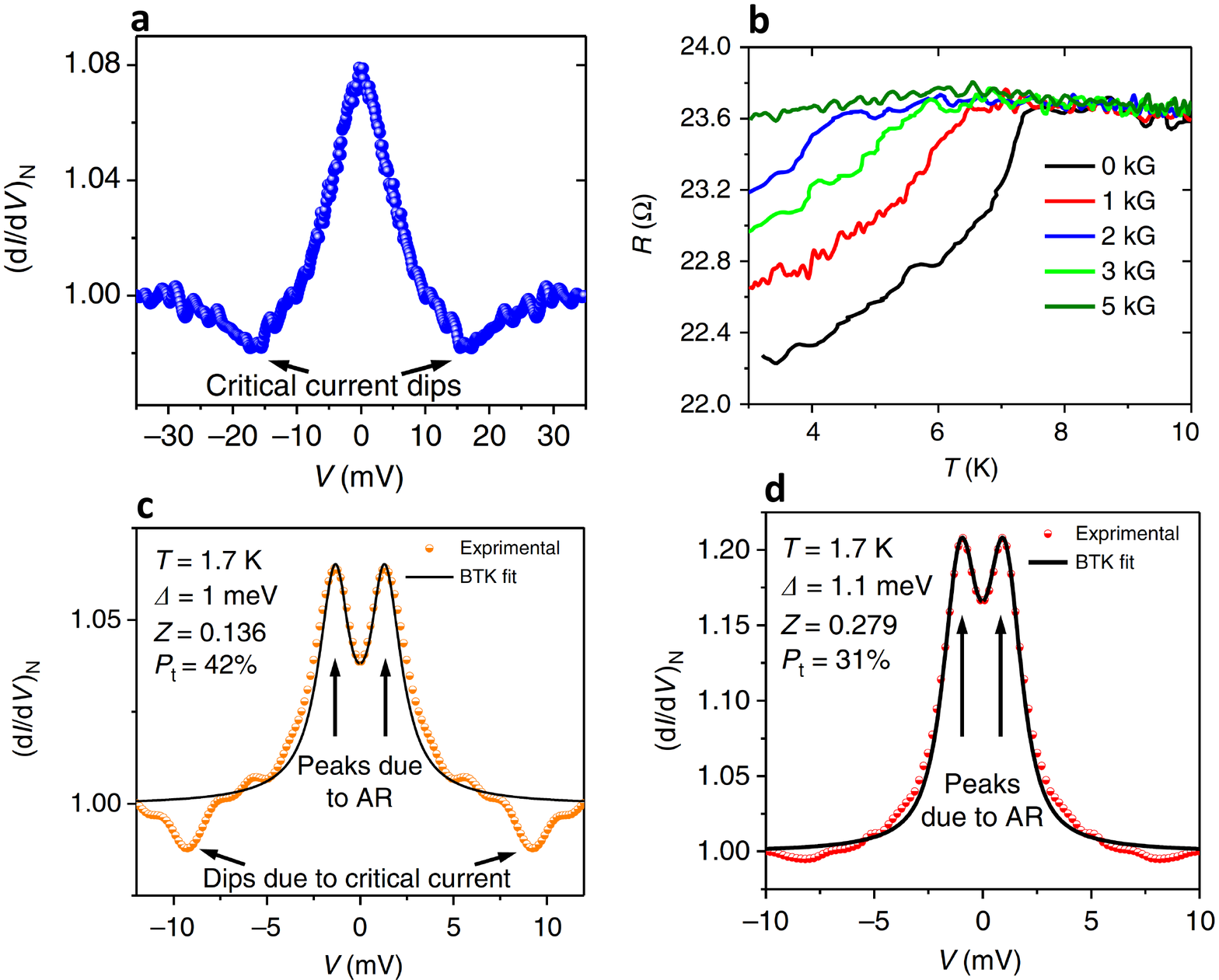}
	\caption{(a) A representative TaAs/Ag spectrum obtained in the thermal limit of transport at 1.8 K,(b) H-dependence of point contact R-T showing a transition at 7K that mimics a superconducting transition. (c) A spectrum obtained in the intermediate regime of transport showing the characteristic signatures of critical current and AR. (d) A ballistic limit experimental spectrum showing only Andreev reflection peaks.}
	\label{int}
\end{figure}

The presence of TISC phase was confirmed for Ag/TaAs point contact in the intermediate and ballistic regimes of mesoscopic transport. In fig. 20(c) a spectrum obtained in the intermediate regime of transport is shown where in addition to the critical current dominated dips in conductance, two peaks symmetric about V=0 corresponding to AR are observed. The Ag/TaAs point-contact was driven to ballistic/diffusive regime of transport by reducing the contact diameter where the double peak structure associated with AR is clearly seen(fig.20(d)).\\
\newpage
\textbf{a. Temperature and magnetic field dependence} From the data presented above it is clear that Ag/TaAS point-contacts exhibit superconducting phase. Now to get a complete picture, it is important to focus on the nature of superconductivity arising in the Ag/TaAs point contacts. In this section, data corresponding to the temperature and magnetic field dependent studies conducted on Ag/TaAs point contacts mainly in the ballistic regime of transport(fig.21 (a),(b)). In the figure, the dotted lines represent the experimental data points and the solid lines correspond to the fits within the modified BTK framework.                          In the presented case the experimental data points matched remarkably well with the theoretical fits. This is quite surprising due to superconducting phase being derived from a complex system, Weyl semimetal TaAs. The evolution of energy gap($\Delta$) with temperature, extracted from the data in fig. 21(a), is shown in fig. 21(c) along with the expected $\Delta$ vs. T curve for a BCS superconductor. \\
erconducting energy gap determined through the BTK fits is 1.2 meV which systematically decreased with increasing temperature. However, the systematic evolution of energy gap with temperature do not coincide with the BCS prediction\cite{Naidyuk, Tinkham}. Moreover, the gap disappeared completely at the observed T$_c$ and unlike Cd$_3$As$_2$, any features associated with the pseudogap were absent\cite{Aggarwal2}. All of the spectra were well fitted under the BTK model along with a high spin polarization indicating a contribution of s-wave component in the order parameter. Additionally, it was observed that the temperature dependence of $\Delta$ deviated from the BCS prediction.  The deviation of certain spectra(ballistic/diffusive regime)from the theoretical fits was observed and considering the deviation of $\Delta$ vs. T from BCS, indicates the possibility of a mixed angular momentum symmetry of the order parameter where an unconventional component is mixed with a strong s-wave component\cite{Gonnelli, Tanaka3, Sato4}. However, the possibility of existence of multiple gaps cannot be ruled out and further experimentation is necessary to reach a conclusion\cite{Golubov}. It is notable here that the extraction of spectroscopic information requires the analysis of spectra showing features associated with AR where critical current dips are absent and the normal state resistance remains independent of temperature. This is precisely what has been done in the presented case.\\
 
 \begin{figure}[h!]
	\centering
	\includegraphics[scale=0.5]{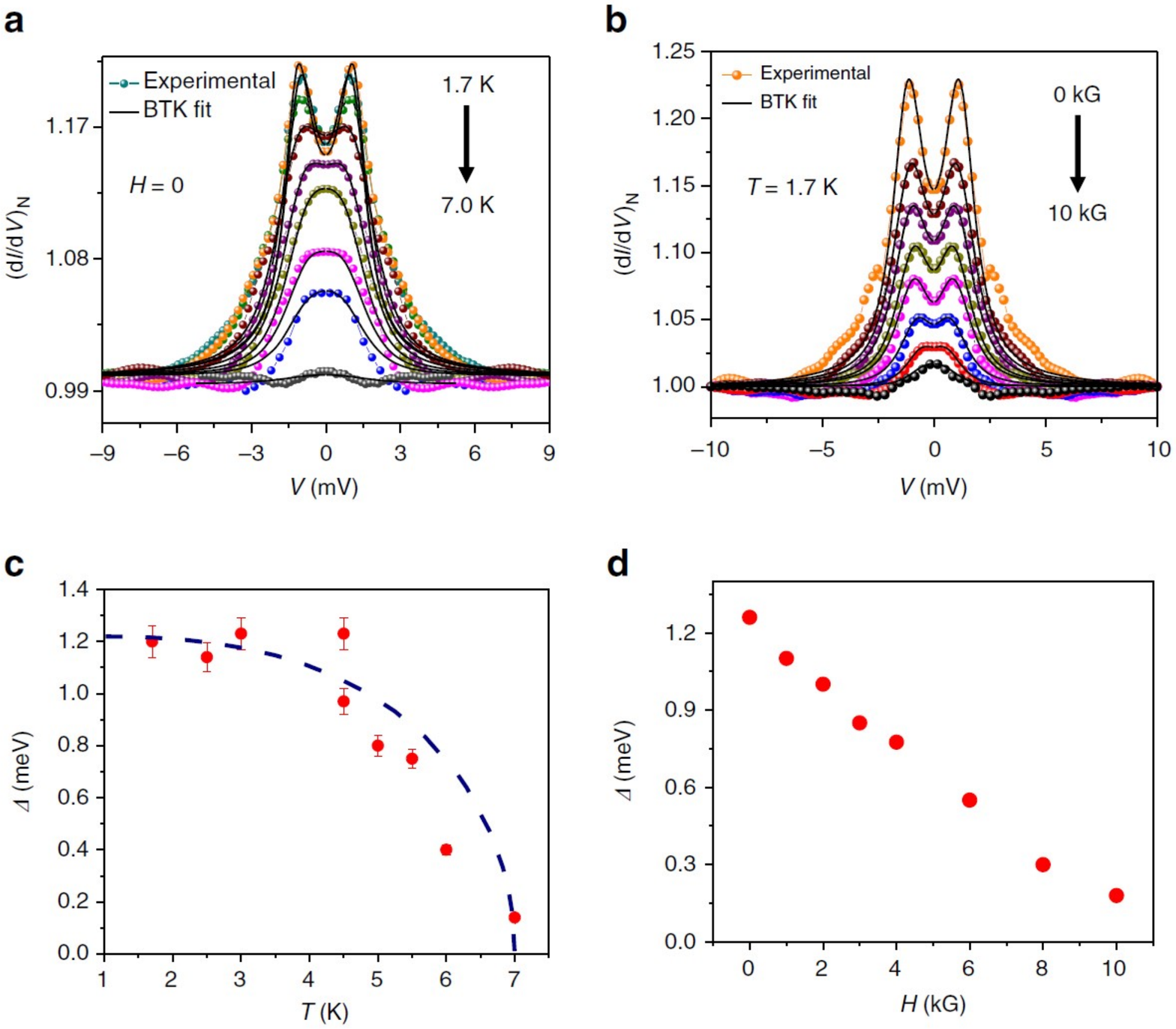}
	\caption{(a)Temperature dependence of the ballistic limit spectra(colored dots) with corresponding theoretical fits(solid black lines), (b) Magnetic field dependence of the ballistic limit spectra(colored dots) along with the T=theoretical fits(solid black lines), (c) Temperature dependence of the gap ($\Delta$). The dashed line shows the expected BCS temperature dependence. The error bars depict the range of $\Delta$ for which a reasonable fit to the experimental spectra could be obtained. (d) H-dependence of the gap ($\Delta$).}
	\label{int}
\end{figure}
 The maximum sup
 Now we focus on the variation of differential conductance spectra with applied magnetic field. The gap structure( double conductance peak in the differential conductance spectrum), as expected, decreases with increasing magnetic field. In the case of a point contact where the contact diameter is large enough to accommodate any vortices, the fitting of AR spectra becomes non-trivial. Moreover, since at present an exact theory describing the TISC remains to be unproven, the presented data cannot be analyzed to confirm whether vortices can enter the point contact region nor if multiple vortices can exist there. All of the field dependent spectra were analysed and fitted under the BTK theory with inclusion of spin polarization. On the basis of analysis and fitting, $\Delta$ vs. H plot was constructed(fig. 21(d)) and it was found that $\Delta$ vanishes at 10 kG.\\
\textbf{b. Spin polarized surface states in TaAs} 

The spectra obtained for Ag/TaAs point contacts, when compared to the expected spectra for a simple elemental superconductor, showed a significant suppression of AR. It is known that the presence of spin polarized surface states causes suppression of Andreev Reflection\cite{Xu6}.

\begin{figure}[h!]
	\centering
\includegraphics[scale=0.7]{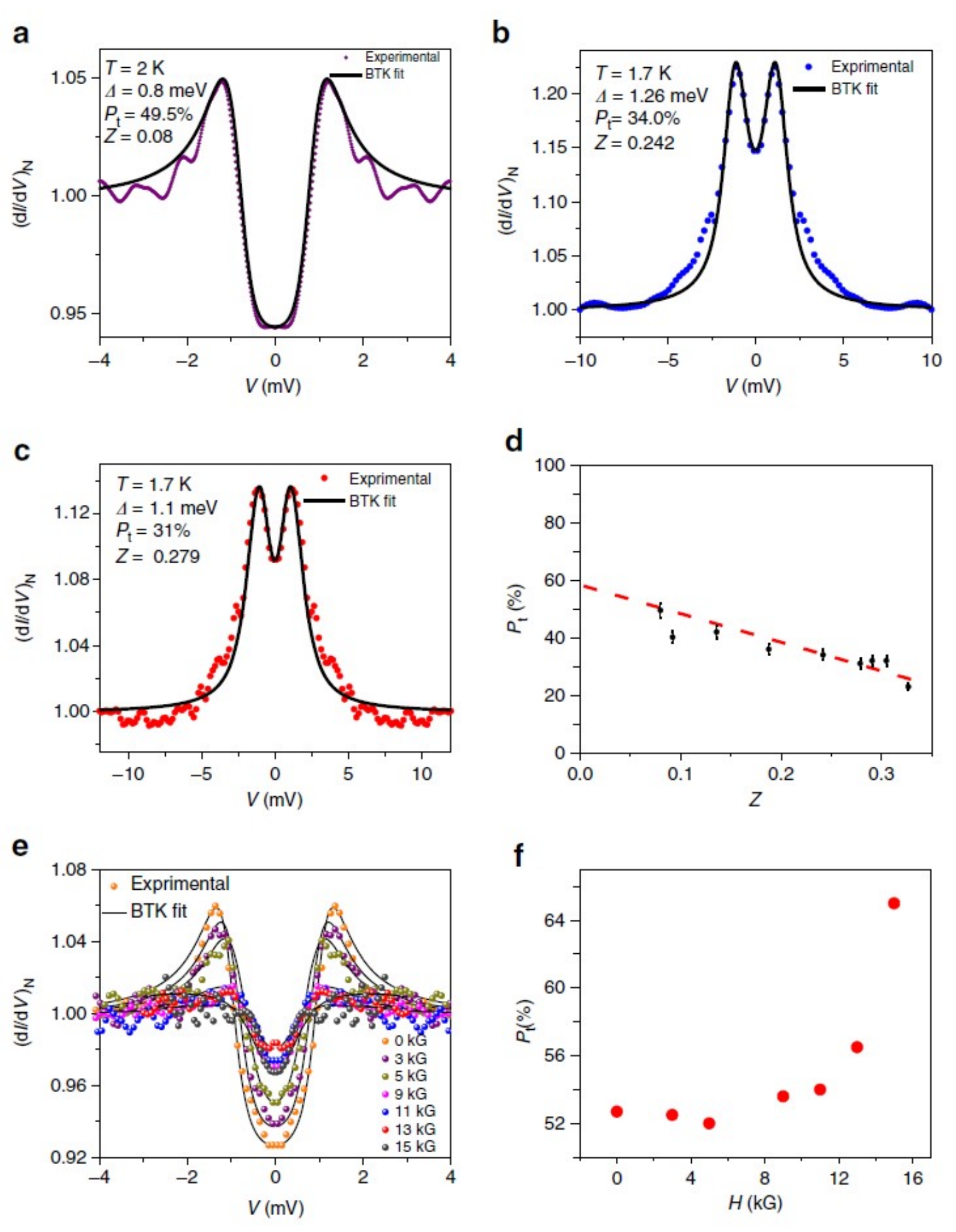}
	\caption{(a),(b),(c) TaAs/Ag spectra(dots) in the ballistic limit and the corresponding theoretical  fits (solid lines) showing strong transport spin polarization. (d) Barrier (Z) dependence of spin polarization. A linear extrapolation of this dependence  to Z=0 shows a large intrinsic transport spin polarization ($\sim$ 60\%). (e) Magnetic field dependence of one of the ballistic point contact spectrum showing the high spin polarization. (f) Magnetic field dependence of spin polarization of the spectrum in (e).}
	\label{int}
\end{figure}

 Such spectra can be fitted within the BTK formalism-modified to include finite transport spin polarization\cite{BTK, Mazin, Soulen}. As mentioned above, all the spectra obtained for the Ag/TaAs point contacts were a remarkable fit within the modified BTK formalism\cite{Plecenik, Sirohi2} and indicated the presence of a large transport spin polarization of about 60\%.  Three such spectra (ballistic/diffusive) with a high spin polarization are illustrated in fig. 22 (a),(b),\&(c). A plot of the detected spin polarization against the barrier potential (Z) is shown in fig. 22(d), where a decrease in transport spin polarization was observed with increasing Z. A linear extrapolation of the curve indicated an intrinsic transport spin polarization of 60 \%. In presence of an external magnetic field, the measured transport spin polarization for a finite Z is seen to increase with increasing magnetic field(fig. 22(e),(f)). Previously spin polarization of TaAs was determined to be 80 \% through ARPES measurement. However, spin polarization determined by PCS is around 60 \% and lower than determined from ARPES measurement. This is due to the fact that PCS is a transport measurement where, the spin polarization of the transport current is measured  rather than the absolute spin polarization\cite{Mazin, Sirohi2}. The presented results and presented analysis indicate the flow of highly spin polarized super-current through the TaAs point contact. 

\textbf{c. Anisotropic magnetoresistance}

Apart from the temperature and field dependent evolution of the contact resistance, a field-angle dependent study of the point contact resistance of a ballistic contact was also explored. Here, the resistance of the contact was measured while the direction of magnetic field is rotated using a 3-axis vector magnet with respect to the direction of injected current. A bias voltage of 13 mV, corresponding to the normal state of TISC, was applied across the contact and magnetic field was rotated.  A large anisotropy in the magneto-resistance was observed which kept increasing with the increse in applied field strength(fig. 23(a). To explain the anisotropy in magneto-resistance, assume that the microconstriction is shaped as a nanowire and the field is rotated with respect to the direction of flow of current through the nanowire. A similar angular magnetoresistance has been previously observed in hybrid nano-structures involving materials whose surface states have complex spin texture\cite{Shang, Kandala2}. A repetition of the experiment at voltage bias (V= 0.3 mV) corresponding to the superconducting state of the TISC(fig.23(b)). The observed magneto-resistance remained anisotropic and equally noticeable as in the previous case. On the basis of these results and observations provide clear indication of coexistence of superconductivity and large spin polarization on TaAs point contacts. 

\begin{figure}[h!]
	\centering
	\includegraphics[scale=0.6]{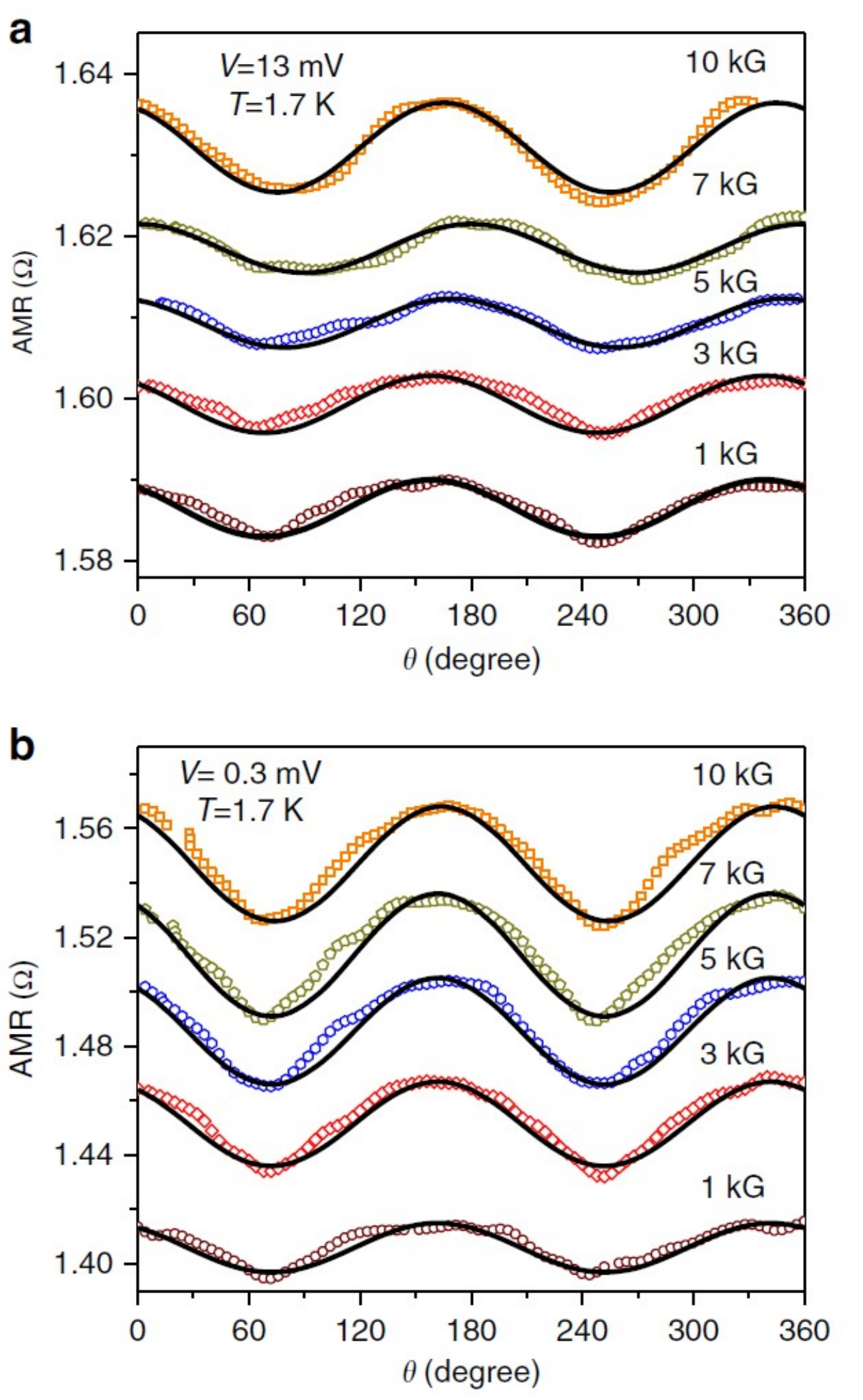}
	\caption{ Anisotropic magnetoresistance of a point contact in the (a) normal state (V=10 mV) and (b) in the superconducting state (V=0.3 mV). The solid black lines are  fits}
	\label{int}
\end{figure}

Further confirmation to the theory that spin polarized current causes emergence of AMR in the Ag/TaAs point contacts is provided from the fact that the AMR data could be fitted well using the typical $cos^2\theta$ dependence. The fits are shown in fig. 23. It is important to consider that the AMR data presented were obtained for a point contact either in the ballistic or diffusive regimes of transport. For such contacts the resistance arises predominantly from the Sharvin's resistance which is independent of the bulk resistance of either of the materials forming the point contact. Hence, it was concluded that the large AMR observed in Ag/TaAs point contacts did not arise due to bulk TaAs. \\
    With the data and discussion presented above, the authors attempt to explain the mechanism causing emergence of TISC phase in Ag/TaAs pont contact. TISC phase may emerge due to local pressure, local doping and/or confinement effects. The TISC phase could not be realized at macroscopic interfaces of metallic silver films deposited on TaAs. Moreover, TaAS is not known to exhibit pressure induced superconductivity. Hence, combination of all three mechanisms mentioned above could be responsible for emergence of TISC on TaAs point contacts. To understand the exact mechanism facilitating the emergence of TISC additional experiments must be performed. One such experiment involves use of double-probe scanning tunnelling microscope capable of determining the exact symmetry of the order parameter. In the double-probe STM the first probe forms the contact with the sample to induce a TISC phase, the second probe is then used to perform spectroscopy in the TISC region i.e close to the first tip attached to probe forming the contact. Such experiments can also probe the possibility of topological and FFLO superconductivity. 

The TISC phase has also been recently achieved on another weyl semimetal TaP\cite{LuO}.

\subsubsection{\textbf{TISC in  nodal semimetals(ZrSiS, TaAs$_2$ NbAs$_2$ etc.)}}

\textbf{ZrSiS:}
ZrSiS is an abundant, non-toxic and highly stable material with a bandstructure that hosts several Dirac cones forming a Fermi surface with diamond shaped line of Dirac nodes\cite{Sankar, Neupane, Schoop}. The recently discovered Dirac semimetals exhibited  Dirac cones with dispersion linear upto a small energy range, for example, for Cd$_3$As$_2$ the linearity exists only upto $\sim$200 meV. However for ZrSiS the linearity extends upto unusually high value of approximately 2 eV\cite{Schoop, Jin, Ko} making ZrSiS a fascinating topological system where lies an enormous possibility of finding anomalous properties such as large magnetoresistance\cite{Lv4} and quantum oscillations that survive even at  high temperatures\cite{Matusiak}. \\

In the following section we present and discuss in detail, the emergence of a superconducting phase induced upon forming a contact between elemental silver (Ag) on crystalline ZrSiS. The Ag/ZrSiS point contacts showed a transition from normal resistance state to the superconducting state at 7.5 K with a superconducting energy gap determined from PCS to be approximately 1 meV. ZrSiS exhibits a liner band dispersion over a large energy range resulting in a robust topological character against perturbation effects such as carrier doping,  variation in stoichiometry etc. and hence superconductivity in ZrSiS point contacts is an important discovery. Moreover, this implies that the topological properties of ZrSiS cannot be destroyed due to proximity to metallic tip forming the point contact and result in emergence of a superconducting phase i.e. the superconducting phase did not emerge at the expense of the topological nature of ZrSiS. Considering the facts presented above ZrSiS makes a promising candidate for realising topological superconductor. 
\newpage
\textbf{Point-contact spectroscopy in different transport regimes on ZrSiS}

\begin{figure}[h!]
	\centering
\includegraphics[scale=0.7]{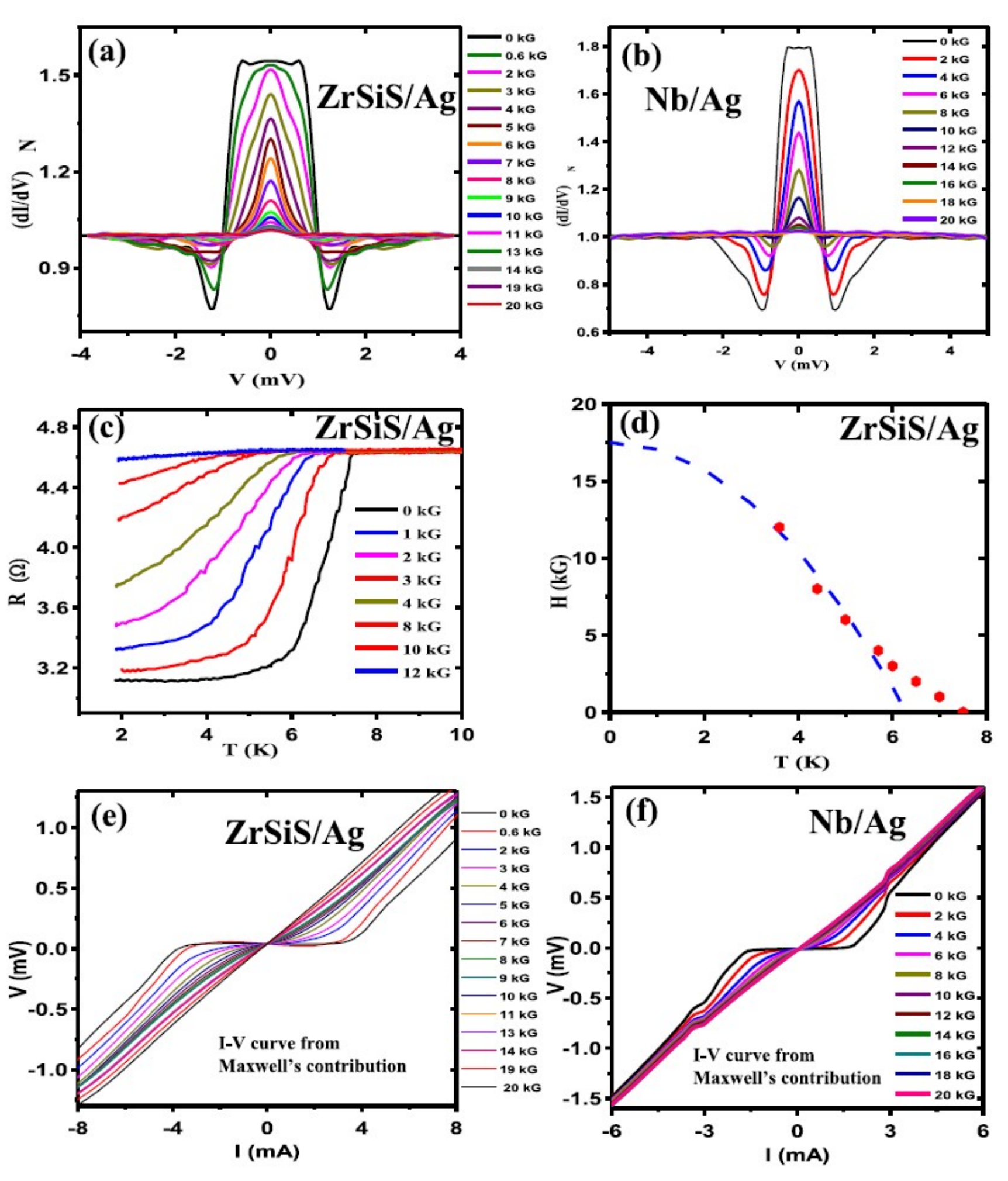}
	\caption{ (a) Magnetic field dependence of spectra obtained in the thermal regime on ZrSiS/Ag point-contact. (b) Magnetic field dependence of spectra in thermal regime on Nb/Ag point-contact. (c) Magnetic field dependent R-T curves showing disappearance of superconducting transition temperature with tincreasing magnetic field. (d) The dashed blue lines represent the H-T phase diagram extracted from R-T curves shown in figure (c). The red line is the expected empirical H-T phase diagram. (e) Evolution of I-V  curves with the magnetic field for Nb/Ag point-contact with only Maxwell's contribution showing  zero resistance portion and evolution with magnetic field. (f) Evolution of I-V  curves with magnetic field for ZrSiS/Ag point-contact with only Maxwell's contribution showing zero resistance  part. }
	\label{int}
\end{figure}

In fig. 24 (a), the thermal regime PCS spectra obtained for ZrSiS/Ag point contact is shown. A striking resemblance is observed with  PCS spectra obtained for known conventional superconductor Nb (fig. 24(b)) with silver(Ag) as tip material. Therefore, it can be inferred that a TISC phase was obtained on ZrSiS/Ag point contact. The spectrum clearly shows critical current dominated conductance dips. Moreover, the spectrum was observed to evolve monotonically with increasing magnetic field. The evolution of contact resistance with temperature is shown in fig. 24(c) where a clear superconducting transition is observed at 7.6 K. Furthermore, as expected for superconductors, the critical temperature decreased monotonically with increasing magnetic field. A H-T phase diagram was constructed from the field dependent R-T curves and is presented in fig. 24(d). The expected empirical H-T phase diagram for a conventional superconductor is shown in fig. 23(d)(blue dotted curve). From the H-T phase diagram it is predicted that the upper critical field for the TISC phase could be as high as 17 kG and is evident from the experiments where,  the experimentally measured data points deviate slightly from the empirical expectation. Moreover, this deviation indicates the existence of an unconventional component in the TISC phase obtained for non-trivial  ZrSiS point contacts.  \\

The authors using the BTK theory, simulated the I-V characteristics associated with the ballistic component(Sharvin's resistance part) of the point contact resistance. The zero resistance state was examined by subtracting the I-V of Sharvin's resistance part from the I-V  curve associated with the  total point-contact resistance, i.e  the I-V curve associated with the Maxwell contribution was obtained and analysed. The I-V curves and corresponding variation with magnetic field is shown in fig. 24(e) and (f) for ZrSiS/Ag and Nb/Ag point contacts respectively. Temperature dependence of the PCS spectra for ZrSiS/Ag point contacts is shown in fig. 25(a),(b) and for Nb/Ag is shown in fig. 25(c),(d). Upon comparison it is confirmed that TISC phase was indeed realized on ZrSiS/Ag point contacts.

\begin{figure}[htb]
	\centering
	\includegraphics[scale=0.5]{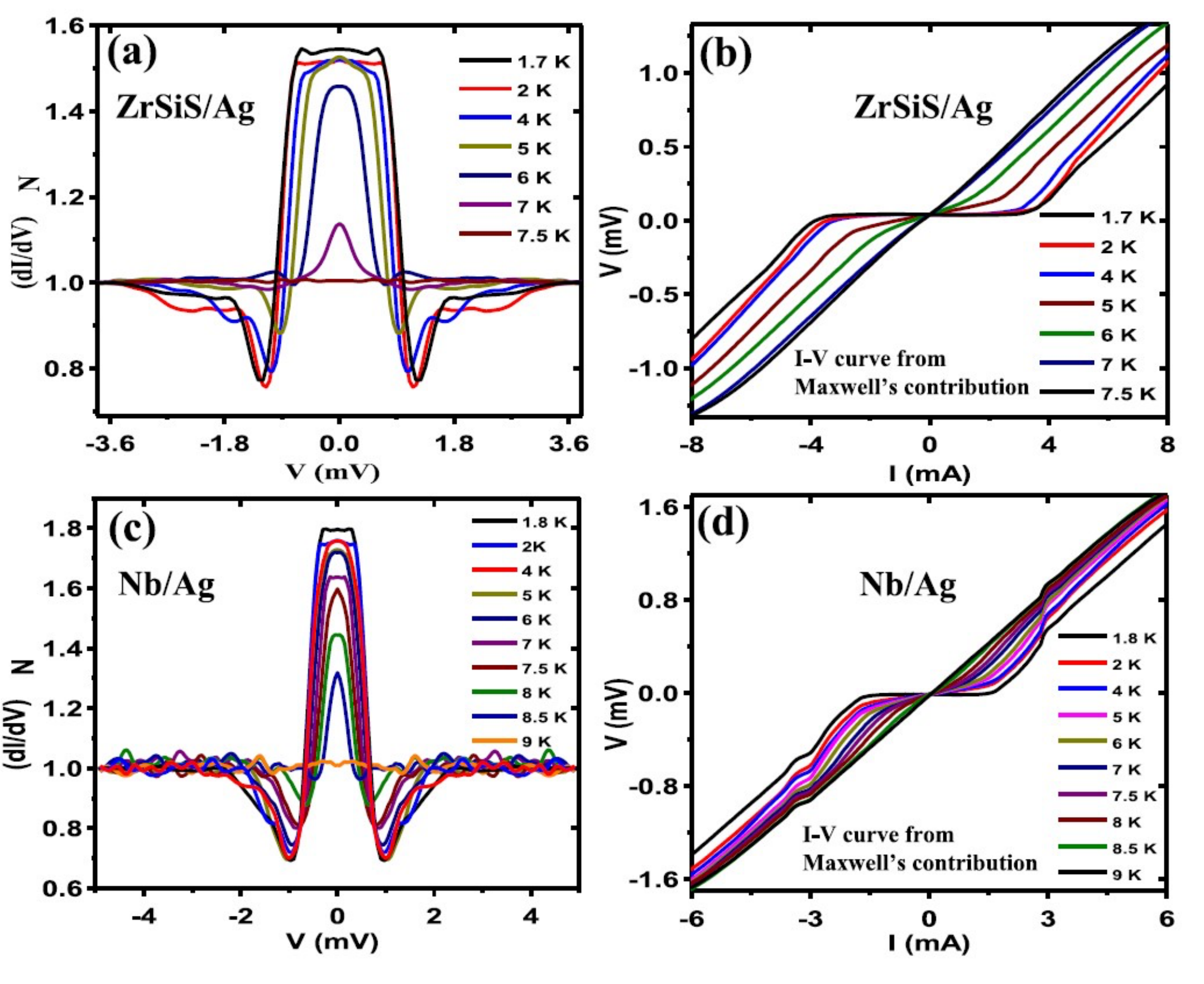}
	\caption{(a) Temperature dependence of thermal spectra obtained for ZrSiS/Ag point-contact. (b) Evolution of I-V curves with temperature for purely Maxwell's contribution  showing zero resistance state in ZrSiS/Ag point contacts. (c) Temperature dependence of the thermal spectra on Nb/Ag piont-contact (d) Evolution of I-V of  Maxwell's contribution to resistance for Nb/Ag point contacts.}
	\label{int}
\end{figure}

On the basis of Wexler's formula, for a point contact in thermal regime, it is possible that a small but finite ballistic component may exist.  A visual inspection of the thermal spectra reveals that the zero field spectra  remains flat between $\pm$0.8 meV. This could be a signature of Andreev reflection for a contact where the barrier potential is extremely small(transparent barrier) and is consistent with BTK prediction. Hence, an estimation of approximated gap amplitude is possible for ZrSiS/Ag point contact. The approximated gap value is $\sim$0.8 meV for which the value of $2\Delta/K_BT_c$ is 2.48. A value consistent with a weak-coupled superconducting phase emerging as a result of phonon mediated pairing of electrons. A first principles calculations of the electronic structure and electron-phonon coupling revealed an enormous increase in DOS at the Fermi level of ZrSiS arising due to formation of contact with metallic tips(fig. 26(a), (b)). However, a considerably smaller contribution of Silver in terms of DOS at and around Fermi level indicated that a substantial increase in carrier density around E$_F$ played an explicit role in emergence of TISC phase in ZrSiS point contacts. \\

\begin{figure}[htb]
	\centering
	\includegraphics[scale=0.6]{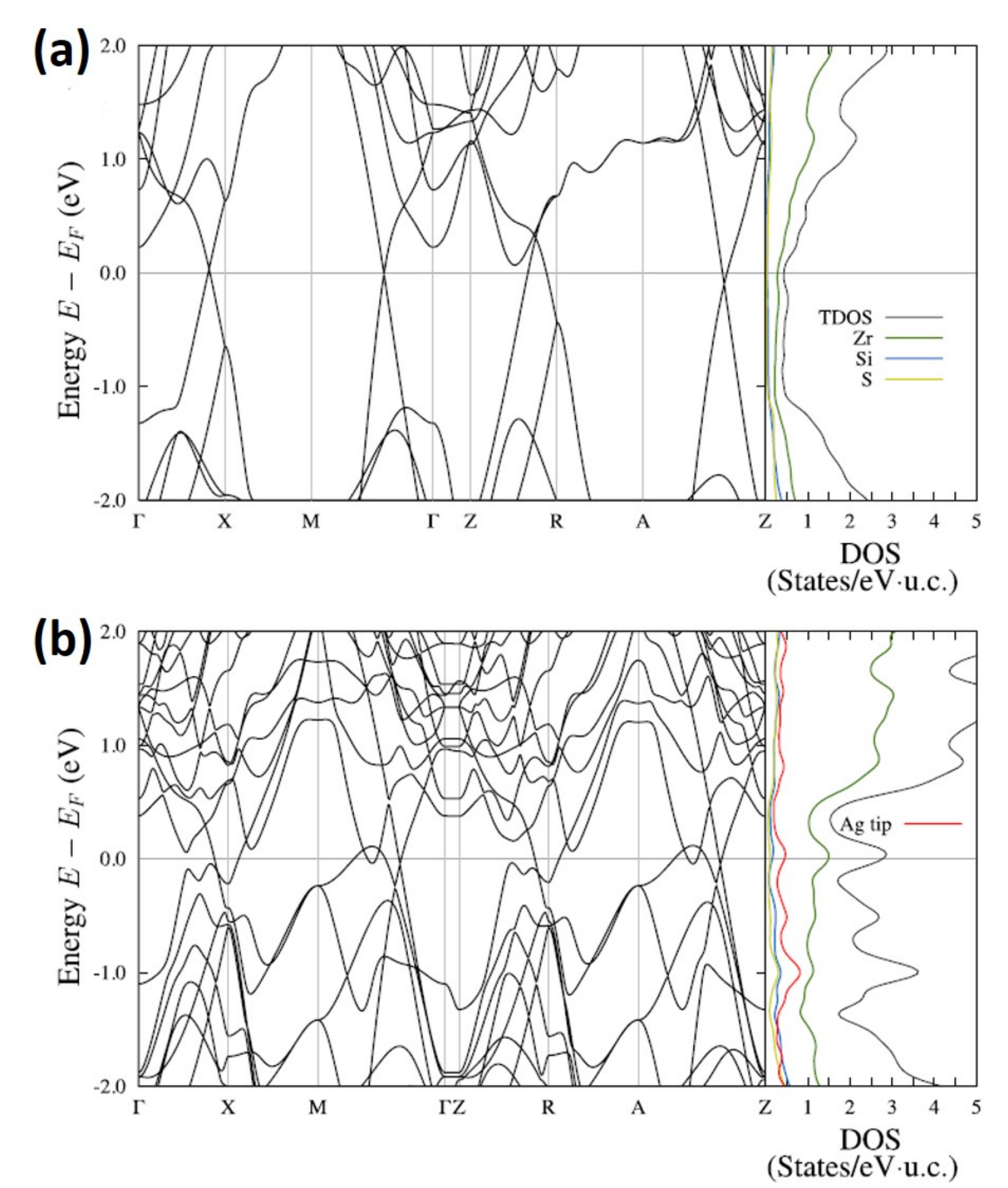}
	\caption{(a)Calculated band structure of ZrSiS showing Dirac cones along $\Gamma$X, M$\Gamma$ and AZ lines of Brillouin zone and corresponding DOS around the Fermi level. (b) Bandstructure of ZrSiS/Ag point contact showing the Dirac cones along M$\Gamma$ and AZ lines with corresponding DOS profile around the Fermi level. }
	\label{int}
\end{figure}
 The electron-phonon coupling was also calculated for bulk ZrSiS and ZrSiS/Ag point contacts. The calculation correlated with the DOS at Fermi energy and it was observed that for ZrSiS formation of microscopic contact with silver leads to an almost two folds increase in $\Lambda$ parameter than the bulk. However, the calculated T$_C$ was two orders of magnitude smaller than the experimental value. Hence it was argued that conventional phonon mediated cooper pairing could not give rise to superconductivity. Moreover, the Dirac cone along $\Gamma$X disappeared upon formation of contact while the Dirac cones along M$\Gamma$ and AZ lines were found to be protected along with appearance of extra nesting features(fig. 26(b)). Therefore, it is possible that  topological character of ZrSiS co-exists with TISC and provides strong indication towards the possibility of topological superconductivity at ZrSiS/Ag point contacts. The TISC phase was recently achieved on other nodal semimetals like TaAs$_2$ and NbAs$_2$\cite{Leishan}.
 
 \subsubsection{\textbf{TISC in Topological crystalline insulators(Pb$_{0.6}$Sn$_{0.4}$Te)}}
 
Topological materials like topological insulators(TI) and  Topological crystalline insulators can be driven into novel phases of matter by introduction of perturbations in form of magnetic dopant, disorders, structural distortions etc.\cite{Peng, Kirzhner, Sasakisan, Hor, Hasani, XQi, TSato, YAndo, LFu, JLiu, Zeljkovic, Liang3, Mitrofanov2, Zeljkovic1, Okada1,  Erickson} Using this idea, physicists worked tirelessly to achieve a topological superconducting phase in these materials. It is believed that such exotic superconducting phases may lead to detection of  the elusive Majorana fermions\cite{Leijnse, LFu1, Beenakker5}. Cu intercalated Bi$_2$Se$_3$, being a candidate material for topological superconductivity, has been extensively investigated in recent times\cite{Wray}. One of the most compelling reasons for such a huge interest is the existence of spin-momentum locking in this system which results in  superconductivity. \\
The protection mechanism of surface states in TCI's is fundamentally different from TIs . The surface states are protected by crystal symmetry in the case of TCI's\cite{LFu, Liang3} whereas,  TIs have time reversal symmetry protected surface states. TCIs possess highly tunable surface states , as compared to TIs, which can be tuned by introduction of perturbations\cite{YAndo}.  Recently, Angle resolved photo-emission spectroscopy confirmed(ARPES) Pb$_{0.6}$Sn$_{0.4}$Te is a TCI. The authors (Shekar et. al) demonstrated that a mesoscopic contact between Pb$_{0.6}$Sn$_{0.4}$Te and an elemental metal  exhibits conventional superconductivity. Moreover, the local superconducting phase was found to exhibit a contact dependent high transition temperature in the range of 3.7 k to 6.5 K.  In the following sections, we discuss the experimental observations in detail. \\
Point contact spectroscopic investigation was performed on Pb$_{0.6}$Sn$_{0.4}$Te using elemental metallic tips, Silver (Ag) and Palladium(Pd). A differential conductance(dI/dV) was measured for the point contacts against an applied bias voltage (V) using a lock-in based modulation technique(Fig. 27(c),(d)).   The resulting spectra showed two sharp conductance dips  which is similar to the conductance features obtained for Pb/Ag and Cu/Nb point contacts(Fig. 27(a),(b)). These dips appear due to the critical current of superconductor. Additionally a zero bias peak also appears in the conductance spectrum which  indicates that the superconducting part of the point contact undergoes a transition to the zero resistance state\cite{GSheet}. The inset of Fig. 27(c) and 27(d) shows the respective resistivity curves showing the transition  temperatures of 5.5 K and 6.3 K for the Ag/ Pb$_{0.6}$Sn$_{0.4}$Te and Pd/ Pb$_{0.6}$Sn$_{0.4}$Te point contacts. The authors observed such critical current dominated features in dI/dV spectrum as well as a transition in the resistivity for a number of different point contacts. Hence, it was inferred that the point contacts on Pb$_{0.6}$Sn$_{0.4}$Te are superconducting. However, to affirm this claim one needs to provide additional data that validates superconductivity. \\

\begin{figure}[htb]
	\centering
	\includegraphics[scale=0.5]{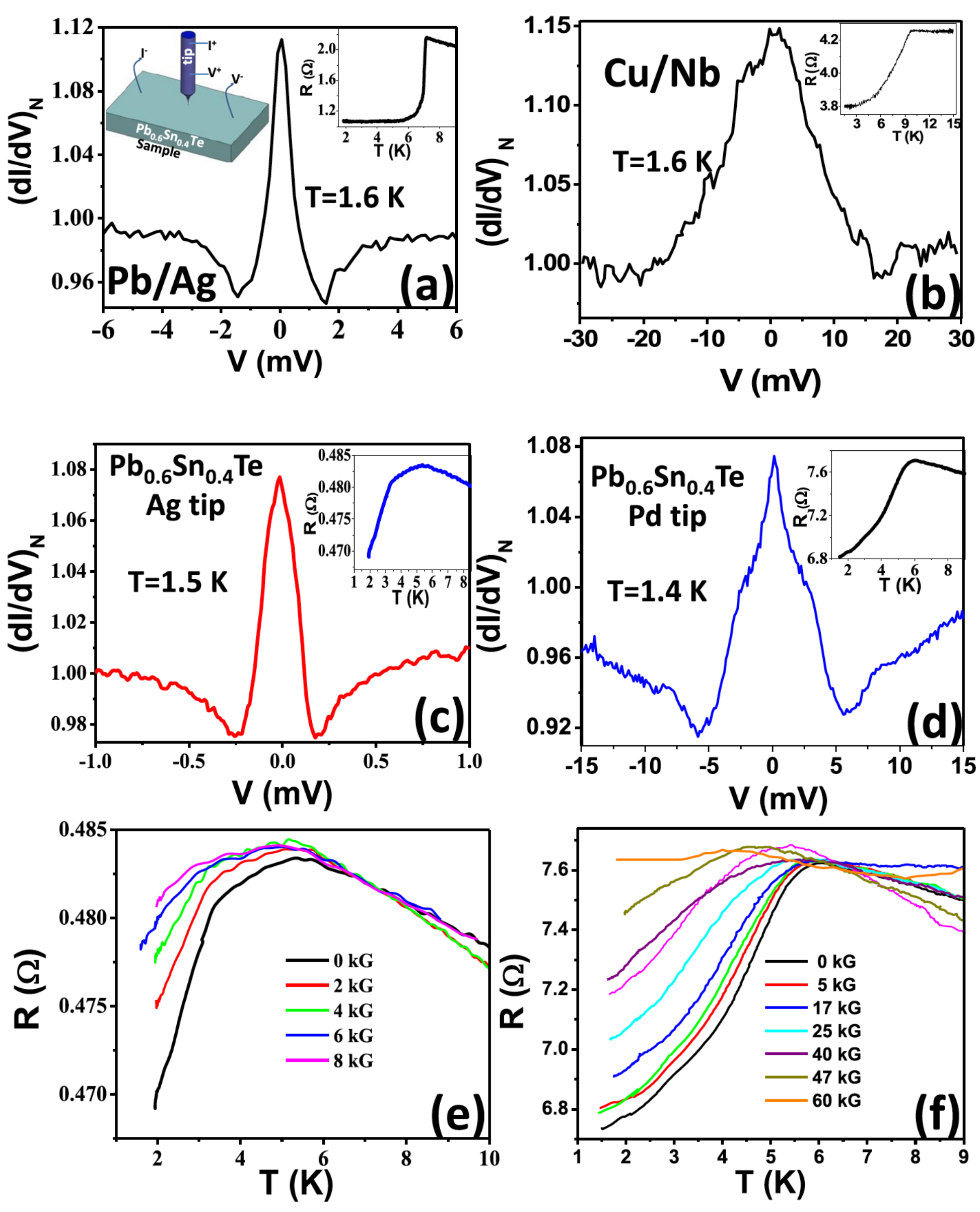}
	\caption{ A differential conductance spectra obtained for (a) Pb/Ag (b) Cu/Nb (c) Pb$_{0.6}$Sn$_{0.4}$Te/Ag (d) Pb$_{0.6}$Sn$_{0.4}$Te/Pd point contact in the thermal regime with corresponding R-T measurement shown in inset. (e),(f)  Variation of R-T with magnetic field for Pb$_{0.6}$Sn$_{0.4}$Te/Ag and Pb$_{0.6}$Sn$_{0.4}$Te/Pd point contacts respectively.}
	\label{int}
\end{figure}

\begin{figure}[htb]
	\centering
	\includegraphics[scale=0.4]{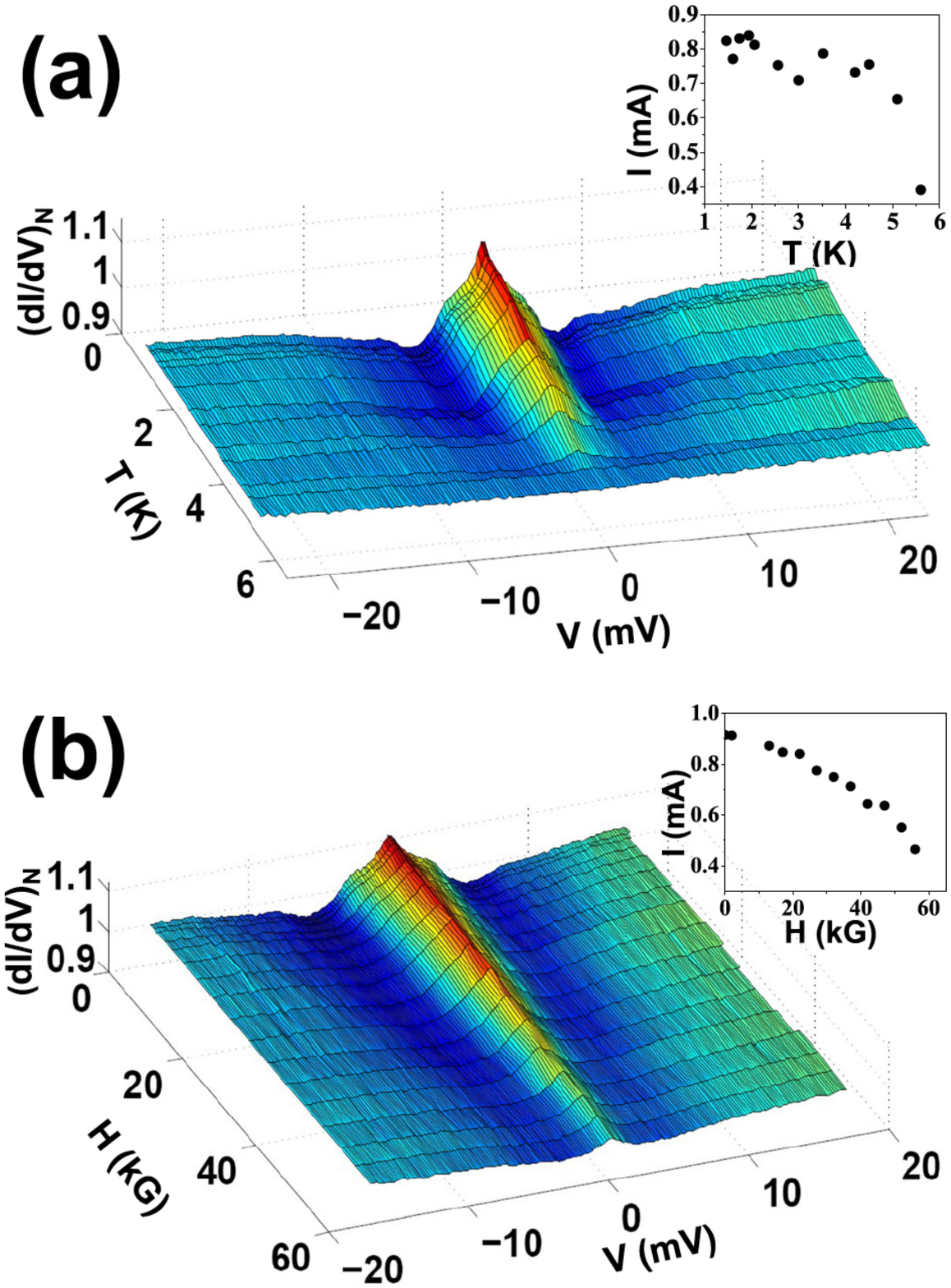}
	\caption{(a)Evolution of differential conductance spectrum of a Pb$_{0.6}$Sn$_{0.4}$Te/Pd point contact with temperature. The inset shows the temperature dependence of critical current. (b) Evolution of differential conductance spectrum of a Pb$_{0.6}$Sn$_{0.4}$Te/Pd point contact with applied magnetic field. The inset shows the field evolution of critical current I$_C$ }
	\label{int}
\end{figure}

The authors presented additional data and arguments supporting the three key aspects pertaining to superconductivity discussed below.\\
1) R-T data showing  systematic field dependence : In Fig. 27(e) and 27(f), the authors presented the Resistance vs Temperature characteristics, for the two point contacts formed on Pb$_{0.6}$Sn$_{0.4}$ with Ag and Pd tips respectively, in presence of various applied magnetic fields.  A clear systematic evolution of superconducting transition temperature with applied magnetic field was observed. The authors observed that the transition temperature systematically decreased with increasing magnetic field, which is an exact behaviour expected for a superconductor. As discussed in previous sections, we remind you that though the overall resistance is zero for a bulk superconductor, in a point contact geometry the resistance is always finite. This is evident from the R-T data and corresponding differential conductance spectrum obtained for Pb/Ag point contact, presented in Fig. 27 (a). \\
2)Systematic temperature dependence of dI/dV spectra:   In Fig. 28(a), We present the temperature dependence of differential conductance spectrum obtained for Pd/Pb$_{0.6}$Sn$_{0.4}$. A cursory look on the data reflects that the dI/dV spectrum smoothly evolves with temperature until they disappear completely at the critical temperature (6 K). Additionally, the critical current of the point contact can be estimated from the position of the conductance dips in the dI/dV. The estimated critical current is presented as a function of temperature in the inset of the Fig. 28(a).  It is also clearly seen that the critical current decreases with increase in temperature. This temperature driven behaviour of critical current is consistent and expected for a superconducting point contact. Moreover, the contact diameter was estimated to be around 50 nm from the normal state resistance via use of Wexler's Formula\cite{Wexler}. \\
3) Systematic field evolution of dI/dV spectra and disappearance at H$_c$: In Fig. 28(b), the field evolution of differential conductance spectra is presented. Here, it is clear from a visual inspection that the dI/dV spectra evolves smoothly with increasing magnetic field where the spectral features diminish monotonically and the spectral features (peaks and dips) completely vanish at a critical field(H$_c$) of 5 T. The field dependence of critical current is presented in the inset of Fig. 28(b).

\begin{figure}[htb]
	\centering
	\includegraphics[scale=0.6]{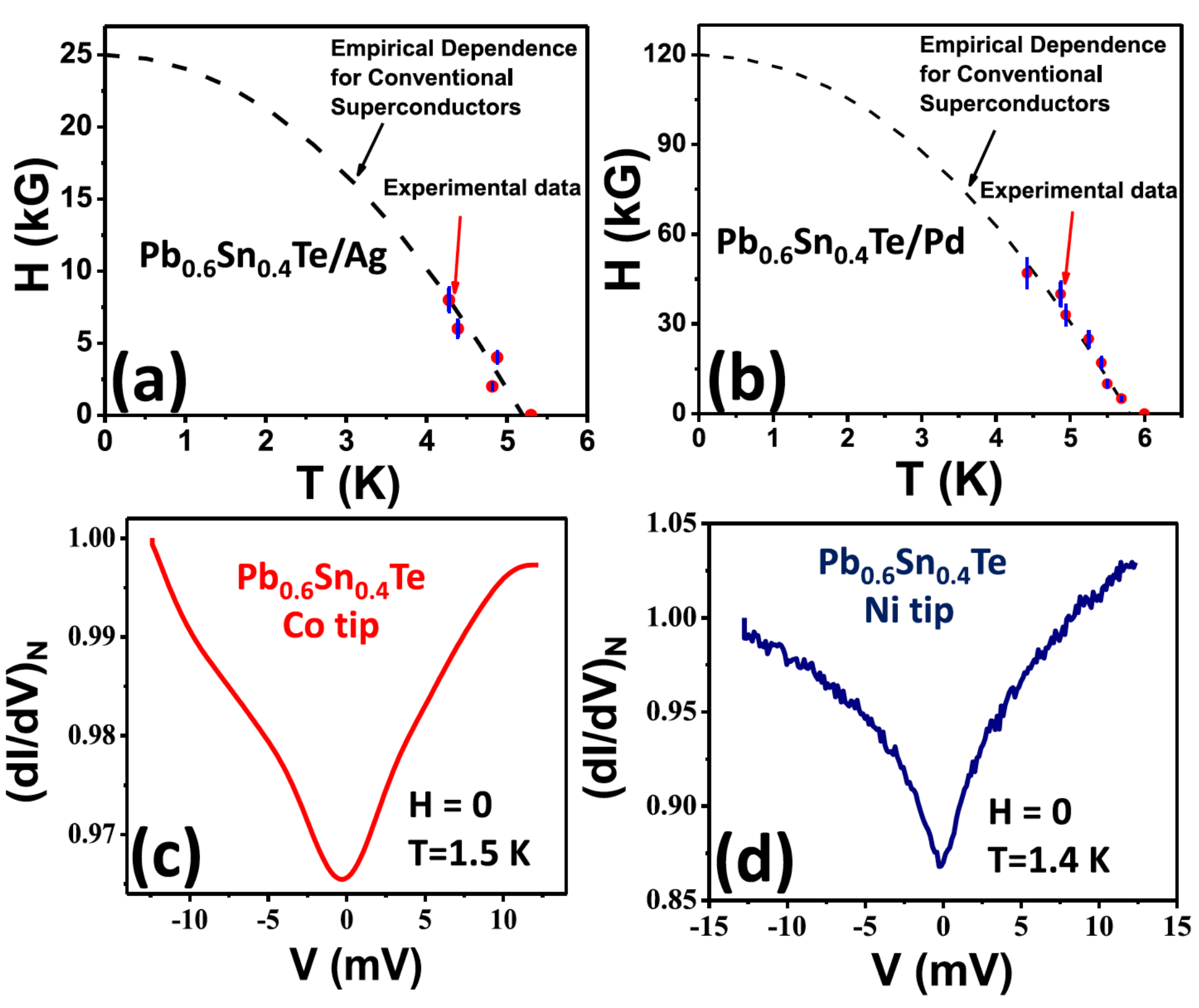}
	\caption{H-T phase diagram for (a) Pb$_{0.6}$Sn$_{0.4}$Te/Ag and (b) Pb$_{0.6}$Sn$_{0.4}$Te/Pd point contact with curve showing the empirical dependence for conventional superconductors. (c),(d) A differential conductance spectrum obtained for point contacts on Pb$_{0.6}$Sn$_{0.4}$Te with  Cobalt(Co) and Niobium(Nb) respectively.\\
	}
	\label{int}
\end{figure}
These results confirmed the existence of a superconducting phase in Pd/Pb$_{0.6}$Sn$_{0.4}$ point contacts. Now the authors focused on determining the nature of superconducting phase in these point contacts for which they constructed that H-T phase diagram. The H-T phase diagram is presented in Fig. 29(a) and 29(b) for Ag/Pb$_{0.6}$Sn$_{0.4}$ and Pd/Pb$_{0.6}$Sn$_{0.4}$ point contacts respectively. The experimentally obtained H-T phase curves were observed to fall on the empirically expected H-T phase curve for conventional superconductors. This indicates that these point contacts exhibit a conventional superconducting phase. 
Pb$_{0.6}$Sn$_{0.4}$ is a topologically non-trivial material where we the authors realized a superconducting phase,  hence, it is important to address  the symmetry of superconducting order parameter where a p-wave symmetry may be possible\cite{LFu1}. This is infact due to the natural breaking of time reversal symmetry, result of p- wave symmetry, where superconducting behaviour is favoured due to proximity of a spin-polarized Fermi surface. This is at odds with  conventional superconductors where proximity to a ferromagnet results in competition between the ferromagnetic order and superconducting order and in turn leads to suppression of the superconducting order. \\

The authors performed point contact spectroscopic investigation on Pb$_{0.6}$Sn$_{0.4}$ using ferromagnetic tips Co and Ni presented in Fig. 29(c) and Fig. 29(d) respectively. In both of the cases, the dI/dV spectra showed a sharp dip in conductance at zero-bias signifying a gap in the density of states while signature of Andreev reflection and/or critical current dominated features were absent. 

Therefore, in this review article, we discussed point contact Andreev reflection spectroscopy in different regimes of transport. We highlighted how such features can be used to detect and confirm the existence of a superconducting phase, called tip-induced superconductivity (TISC), that appears only under a mesoscopic point contact between two non-superconducting materials, most of the times one of these being a topologically non-trivial system.

\section{Acknowledgements}
The authors thank  Leena Aggarwal, Sirshendu Gayen, Soumya Datta  and Ritesh Kumar.

\end{document}